\documentclass{aastex}
\usepackage{spr-astr-addons}             
\usepackage{url}
\usepackage[utf8x]{inputenc}
\usepackage{lscape, morefloats, graphicx, float, soul}   
\RequirePackage{color}                   

\begin{document}

\title{A study of the Czernik 2 and NGC 7654 open clusters using CCD UBV photometric and Gaia EDR3 data}
\slugcomment{Not to appear in Nonlearned J., 45.}
\shorttitle{A study of the Czernik 2 and NGC 7654 open clusters}
\shortauthors{B. Akbulut, S. Ak, T. Yontan, Bilir, S., T. Ak, T. Banks, E. Kaan Ulgen, E. Paunzen}

\author{B. Akbulut \altaffilmark{1}}
\altaffiltext{1}{Istanbul University, Institute of Graduate 
Studies in Science, Programme of Astronomy and 
Space Sciences, 34116, Beyaz{\i}t, Istanbul, Turkey\\
Corresponding Author: burcu.akbulut@ogr.iu.edu.tr\\}

\author{S. Ak \altaffilmark{2}}
\altaffiltext{2}{Istanbul University, Faculty of Science, Department 
of Astronomy and Space Sciences, 34119 University, Istanbul, Turkey\\}

\author{T. Yontan \altaffilmark{2}}
\altaffiltext{2}{Istanbul University, Faculty of Science, Department 
of Astronomy and Space Sciences, 34119 University, Istanbul, Turkey\\}

\author{S. Bilir \altaffilmark{2}} 
\altaffiltext{2}{Istanbul University, Faculty of Science, Department 
of Astronomy and Space Sciences, 34119 University, Istanbul, Turkey\\}

\author{T. Ak\altaffilmark{2}} 
\altaffiltext{2}{Istanbul University, Faculty of Science, Department 
of Astronomy and Space Sciences, 34119 University, Istanbul, Turkey\\}

\author{T. Banks\altaffilmark{3, 4}} 
\altaffiltext{3}{Nielsen, Data Science, 200 W Jackson Blvd \#17, Chicago, IL 60606, USA\\}
\altaffiltext{4}{Physics \& Astronomy, Harper College, 1200 W Algonquin Rd, Palatine, IL 60067, USA\\}

\author{E. Kaan Ulgen\altaffilmark{1}}
\altaffiltext{1}{Istanbul University, Institute of Graduate 
Studies in Science, Programme of Astronomy and 
Space Sciences, 34116, Beyaz{\i}t, Istanbul, Turkey\\}
\and
\author{E. Paunzen \altaffilmark{5}}
\altaffiltext{5}{Department of Theoretical Physics and Astrophysics, 
Masaryk University, Kotl\'a\u rsk\'a 2, 611 37 Brno, Czech Republic\\}

\begin{abstract} 
We analysed the open clusters Czernik 2 and NGC 7654 using CCD {\em UBV} photometric and {\it Gaia} Early Data Release 3 (EDR3) photometric and astrometric data. Structural parameters of the two clusters were derived, including the physical sizes of Czernik 2 being $r=5^{\prime}$ and NGC 7654 as $8^{\prime}$. We calculated membership probabilities of stars based on their proper motion components as released in the {\it Gaia} EDR3. To identify member stars of the clusters, we used these membership probabilities taking into account location and the impact of binarity on main-sequence stars. We used membership probabilities higher than $P=0.5$ to identify 28 member stars for Czernik 2 and 369 for NGC 7654. The mean proper motion components ($\mu_{\alpha}\cos\delta$, $\mu_{\delta}$) of Czernik 2  were derived as ($-4.03 \pm 0.04$,$-0.99 \pm 0.05$) mas yr$^{-1}$ and for NGC 7654 as ($-1.89 \pm 0.03, -1.20 \pm  0.03$) mas yr$^{-1}$. We estimated colour-excesses and metallicities separately using $(U-B) \times (B-V)$ two-colour diagrams to derive homogeneously determined parameters. The derived $E(B-V)$ colour excess is $0.46 \pm 0.02$ mag for Czernik 2 and $0.57 \pm 0.04$ mag for NGC 7654. [Fe/H] metallicities were obtained for the first time for both clusters, $-0.08 \pm 0.02$ dex for Czernik 2 and $-0.05 \pm 0.01$ dex for NGC 7654. Keeping the reddening and metallicity as constant quantities, we fitted  PARSEC models using $V\times (B-V)$ and $V\times (U-B)$ colour-magnitude diagrams, resulting in estimated distance moduli and ages of the two clusters. We obtained the distance modulus for Czernik 2 as $12.80 \pm 0.07$ mag and for NGC 7654 as $13.20\pm 0.16$ mag, which coincide with ages of $1.2\pm 0.2$ Gyr and $120\pm 20$ Myr, respectively. The distances to the clusters were calculated using the {\it Gaia} EDR3 trigonometric parallaxes and compared with the literature. We found good agreement between the distances obtained in this study and the literature. Present day mass function slopes for both clusters are comparable with the value of \citet{Salpeter55}, being $X=-1.37\pm 0.24$ for Czernik 2 and $X=-1.39\pm 0.19$ for NGC 7654. 
\end{abstract}


\section{Introduction}           

\begin{table*}[htbp]
\setlength{\tabcolsep}{4pt}
\small
  \centering
  \caption{Basic parameters for Czernik 2 and NGC 7654 derived by this study and compiled from the literature: columns present cluster names, colour excesses $E(B-V$), distance moduli $\mu_V$, distances $d$, iron abundances [Fe/H], age $t$, trigonometric distances from {\it Gaia} $\langle d_{Gaia}\rangle$ and proper motion components $\langle\mu_{\alpha}\cos\delta\rangle$, $\langle\mu_{\delta}\rangle$.}
  \begin{tabular}{ccccccccccc}
    \hline
    Cluster &  $E(B-V)$ & $\mu_V$ & $d$ & [Fe/H] & $t$ &  $\langle d_{\rm Gaia}\rangle$ & $\langle\mu_{\alpha}\cos\delta\rangle$ &  $\langle\mu_{\delta}\rangle$ & Ref \\
            & (mag) & (mag) & (pc)  & (dex) & (Myr) & (pc)  & (mas yr$^{-1}$) & (mas yr$^{-1}$) &  \\
    \hline
    Czernik 2 &  0.74$\pm$0.10 & 13.54$\pm$0.10   & 1775$\pm$80 & ---   & 100 &  ---   & --- & ---  & (1) \\
              &  --- & --- & --- & --- & ---  &  1907$\pm$377 & -4.06$\pm$0.01 & -0.90$\pm$0.01 & (2) \\  
              & --- & --- & --- & --- & 1590$\pm$95 &  2016$\pm$155 & -4.05$\pm$0.02 & -0.96$\pm$0.02 & (3) \\  
              & 0.45 & 12.80 & 1899 & --- & 370 &  2020$\pm$202 & -4.06$\pm$0.01 & -0.90$\pm$0.01 & (4) \\
              & 0.46$\pm$0.02 & 12.80$\pm$0.07 & 1883$\pm$63 & -0.08$\pm$0.02 & 1200$\pm$200 &  1919$\pm$189 & -4.03$\pm$0.04 & -0.99$\pm$0.05 & (5) \\
  \hline
NGC 7654  & --- & --- & --- & --- & ---  & 1600$\pm$263 & -1.94$\pm$0.01 & -1.13$\pm$0.01 & (2) \\ 
          & --- & --- & --- & --- & 380$\pm$23 &  1681$\pm$165 & -1.93$\pm$0.01 & -1.13$\pm$0.01 & (3) \\ 
          
          & 0.60 & 12.94 & 1653 & --- & 155 & 1678$\pm$130 & -1.94$\pm$0.01 & -1.13$\pm$0.01 & (4) \\
          & 0.57$\pm$0.04 & 13.20$\pm$0.16 & 1935$\pm$146 & -0.05$\pm$0.01 & 120$\pm$20 & 1640$\pm$82~ & -1.89$\pm$0.03 & -1.20$\pm$0.03 & (5) \\
          & 0.64 & 12.75 & 1421 & ---   & 58 &  ---   & ---   & ---   & (6) \\
          & 0.66 & 13.04$\pm$0.20   & 1580$\pm$150   & ---   & --- & ---      & ---   & ---   & (7) \\
          & 0.70 & 13.98  & 2300   & ---   & 10 & ---    & ---   & ---   & (8) \\
          & 0.57 &  ---      &  ---      & ---   & 96 &  ---   & ---   & ---   & (9) \\          
          & 0.57 & 12.60  & 1470   & ---   & 35 & ---   & ---   & ---   & (10) \\         
          & 0.58$\pm$0.02 & 12.53$\pm$0.29    & 1400$\pm$200 & ---   & 158 &  ---   & ---   & ---   & (11) \\
          & 0.62$\pm$0.05 & 12.82$\pm$0.20    & 1510$\pm$145 & ---   & 100 &  ---   & ---   & ---   & (12) \\
          & 0.57 & 12.50$\pm$0.10   & 1380$\pm$70 & ---   & 160 &  ---   & ---   & ---   & (13) \\
          & 0.58$\pm$0.03 & 12.60$\pm$0.10   & 1400$\pm$200 & ---   & 60$\pm$10 &  ---   & ---   & ---   & (14) \\
          & 0.73$\pm$0.15 & 13.11$\pm$1.15   & 1480$\pm$470 & --- & 10 &  ---   & ---   & ---   & (15) \\
           \hline
    \end{tabular}%
    \\
(1) \citet{Tadross09}, (2) \citet{Cantat-Gaudin18}, (3) \citet{Liu19}, (4) \citet{Cantat-Gaudin20}, (5) This study, (6) \citet{Lundby46}, (7) \citet{Schmidt77}, (8) \citet{Pfau80}, (9) \citet{Kaltcheva90}, (10) \citet{Battinelli94}, (11) \citet{Viskum97}, (12) \citet{Choi99}, (13) \citet{Pandey01}, (14) \citet{Bonatto06}, (15) \citet{Maciejewski07}

\end{table*}%

\label{sect:intro}

Open star clusters are valuable tools to understand the structure and evolution of the Galaxy. The stars which make up open clusters are formed by the collapse of a molecular cloud under similar physical conditions and subsequently loosely bound to each other by weak mutual gravitational forces. For these reasons the distances, metallicities, and ages of cluster member stars are similar to each other, although their masses differ from star to star. These basic astrophysical properties of open star clusters are important in understanding star formation and evolution.  One of the important findings obtained from the study of open cluster stars having similar ages and chemical structures is that they reflect the physical properties of the cloud they formed from. Cluster stars are a useful `tool' to explore star formation and evolution, stellar interactions,  stellar nucleosynthesis, and the dynamical evolution of the cluster to which they are gravitationally bound.

Open clusters are young stellar groups that consist of around a hundred to several thousand stars. The member stars are gravitationally bound to each other and share the same origin (as explained above). Such clusters represent an important component of the spiral arms of the Milky Way. Also known as Galactic clusters, these objects are rich in gas and dust understood to be the remaining matter from the initial period of star formation. Morphologies vary from cluster to cluster. Open clusters may appear as sparse star distributions and irregularly mixed with field stars, or as crowded, dense, and relatively spherical forms containing several thousand stars. The gravitational potential effects of the Galactic disc can make it difficult to determine the structural properties and internal dynamics of these objects, as well as influencing the life time of the cluster as an observable entity. The level of stellar density affects the determination of structural parameters for clusters: a small open cluster can be dispersed by internal interactions such as the evolution of stars, ejection and evaporation; whereas a large open cluster can be dispersed by external interactions such as giant molecular clouds, the galactic tidal field, and shock waves in the spiral arms \citep{Carraro06, Andersen00}. Since the differences in cluster size also affects their dynamic evolution it is necessary to examine the stellar densities of different radius intervals from the cluster centers to understand the evolution process and determine sensitive parameters.

Photometry, astrometry, and spectroscopy are all methods applied to the analysis of open clusters.  An analysis depending on only one of these methods can lead to results conflicting with an analysis based on another technique; a better approach is to apply as many of these methods together (hereafter called homogenisation) to a given cluster to derive more self-consistent results, such as in this study. Improving our understanding of star formation, stellar evolution, and the structure of the Galaxy requires determining complete spatial, structural, kinematic and astrophysical parameters of many known open clusters. Therefore data of open clusters  have been compiled in catalogues by working groups and continuously updated databases created. Examples of databases and catalogues containing such photometric, spectroscopic and astrometric data of open clusters include WEBDA\footnote{https://webda.physics.muni.cz/} \citep{Mermilliod95}, DAML02\footnote{https://wilton.unifei.edu.br/ocdb/} \citep{Dias02}, 
SAI\footnote{http://ocl.sai.msu.ru/}\citep {Koposov08}, MWSC\footnote{https://heasarc.gsfc.nasa.gov/W3Browse/all/mwsc.html} \citep{Kharchenko12} and UPK\footnote{https://sites.google.com/ushs.hs.kr/upk}\citep{Sim19}. 

In photometric analyses of open clusters, fundamental astrophysical parameters such as colour excesses, metallicity, distance, and age of these objects are determined by using colour-magnitude diagrams (CMDs) and two-colour diagrams (TCDs). The relevant parameters are estimated through the comparison of observational data with stellar models and theoretical isochrones on CMDs and TCDs. Differences in the morphologies of open clusters affect the distributions, densities, and placements of the main sequence, turn off, and giant stars on CMDs and TCDs, which in turn can lead to interpretative issues. Therefore to get reliable results one should take into account only cluster member stars which have been accurately selected, ensure the quality and homogeneity of data used, and employ uniform analysis methods across the clusters being studied. 

The European Space Agency's (ESA) {\it Gaia} Early Data Release 3 \citep[hereafter {\it Gaia} EDR3,][]{Gaia20} supplies updated equatorial ($\alpha$, $\delta$) and Galactic ($l$, $b$) coordinates, trigonometric parallaxes and proper motions ($\varpi$, $\mu_{\alpha}\cos\delta$, $\mu_{\delta}$), and photometric magnitudes ($G$, $G_{\rm BP}$ and $G_{\rm RP}$) of nearly 1.8 billion objects to high accuracies, based on 34 months of observational data. The astrometric data ($\alpha$, $\delta$, $\varpi$, $\mu_{\alpha}\cos\delta$, $\mu_{\delta}$) have a limiting magnitude of $G\sim 21$ mag. At $G\leq20$ mag the uncertainty in the $G$-band photometry ranges across 0.2-6 mmag. For the sources brighter than $G\leq15$ mag, the median error of trigonometric parallaxes and proper motion components are up to 0.03 mas and 0.03 mas yr$^{-1}$, respectively, while these limits reach 1.3 mas and 1.4 mas yr$^{-1}$ up to $G\sim 21$ mag, respectively \citep{Gaia20}. Such accurate data allow us to obtain membership probabilities of stars, mean proper motion components and distance values for the open clusters with high precision.

In this study we determined the structural, astrophysical and astrometric parameters of the Czernik 2 and NGC 7654 open clusters.  We used  {\it Gaia} EDR3 astrometric and photometric data together with CCD {\em UBV} photometric observations, utilising independent methods, presenting detailed {\em UBV} and {\it Gaia} results for both clusters. The literature summaries of the clusters are as follows:

\subsection{Czernik 2}
Czernik 2 ($\alpha=00^{\rm h} 43^{\rm m} 49^{\rm s}$, $\delta=+60^{\rm o} 11^{\rm '} 49^{\rm''}$, $l=121^{\rm o}.98$, $b=-2^{\rm o}.66$) was classified as Trumpler III1r \citep{Ruprecht66} and presented in detail in the literature for the first time by \citet{Czernik66} as an open cluster with an $r=10$ arcmin angular diameter. 128 stars were reported for the cluster in the study. Although this cluster consists of stars with apparent magnitudes to be considered faint, it is classified by \citet{Trumpler30} as easily definable due to its relative greater stellar density compared to the surrounding field stars. The first CCD photometric observations were carried out by \citet{Phelps94}. From an analysis of the CMD, they stated that Czernik 2 did not clearly show characteristics of being an open cluster. \citet{Maciejewski08} used CCD {\em BV} and 2MASS {\em JHK$_S$} observational data to make a photometric and astrometric analysis of the cluster, examining whether Czernik 2 was an open cluster. Analysing together CMDs and vector-point diagrams (VPDs) constructed for different distances, \citet{Maciejewski08} could not define a central concentration in the cluster region nor could they even separate the main sequence of cluster from field stars on the CMDs. For these reasons they claimed that Czernik 2 is not an open cluster. \Citet{Tadross09} examined 60 poorly studied open clusters, including Czernik 2, using {\em JHK$_S$} near-IR 2MASS photometry \citep{Skrutskie06}. They obtained the diameter, reddening, distance and age of Czernik 2 as $r=5.8$ arcmin, $E(B-V)=0.74\pm 0.10$ mag, $d=1775\pm80$ pc and $t=100$ Myr respectively. \citet{Cantat-Gaudin18} made use of {\it Gaia} DR2 data alone to obtain a list of members, mean distance and proper motion components of 1,229 open clusters including new discovered 60 open clusters. They calculated the distance of Czernik 2 from {\it Gaia} DR2 trigonometric parallaxes as $d=1907\pm377$ pc, and mean proper motion components as $\mu_{\alpha}\cos\delta=-4.06\pm0.01$ and $\mu_{\delta}=-0.90\pm0.01$ mas yr$^{-1}$. \citet{Liu19} applied the Star cluster Hunting Pipeline (SHiP) to identify star clusters in {\it Gaia} DR2 data, covering 2,443 open clusters. As a result of their analysis, they determined the distance of the Czernik 2 cluster as $d= 2016\pm155$ pc, its age as $t=1590\pm95$ Myr, and the metal abundance as $\rm \log (Z/Z_{\odot})=-1$. Moreover, they calculated mean proper motion components of Czernik 2 as $\mu_{\alpha}\cos\delta=-4.05\pm0.02$ and $\mu_{\delta}=-0.96\pm0.02$ mas yr$^{-1}$. \citet{Cantat-Gaudin20} put to use the homogeneous {\it Gaia} DR2 photometry within the upper magnitude $G\leq18$ mag to derive the main parameters (distance, age, and interstellar extinction) of 1,867 clusters identified with {\it Gaia} DR2 astrometry. Determinations of the main cluster parameters were done via isochrone fitting, which resulted in the values of distance, age, and extinction of the Czernik 2 being estimated as $d=1899$ pc, $t=370$ Myr, and $A_{\rm V}=1.41$ mag (which corresponds to $E(B-V)=0.45$ mag), respectively.  \citet{Cantat-Gaudin20} also estimated a mean distance $d= 2020\pm202$ pc to the cluster through use of the trigonometric parallaxes of cluster member stars. 

\subsection{NGC 7654}
NGC 7654, also known as M52, ($\alpha=23^{\rm h} 24^{\rm  m} 47^{\rm s}$, $\delta = +61^{\rm o} 35^{\rm '} 36^{\rm''}$, $l=112^{\rm o}.82$, $b=+0^{\rm o}.43$) is an open cluster classified as Trumpler I2r with an angular diameter of 6$'$ and a distinct central stellar density. The cluster has been studied by several researchers who used {\em UBV}, {\em uvby$\beta$} and 2MASS photometries \citep{Lundby46, Pesch60, Schmidt77, Pfau80, Danford81, Kaltcheva90, Battinelli94, Bonatto06}. These studies closely agree on the cluster's astrophysical parameters. The ranges of colour excess and age of NGC 7654 were determined by these different authors as being within $0.57\leq E(B-V)\leq 0.70$ mag and $10\leq t \leq 96$ Myr respectively. The mean distance of the cluster is given as $d=1450$ pc \citep{Lundby46, Pesch60, Schmidt77, Choi99}. \citet{Pandey01} used CCD {\em UBVRI$_c$} photometry to examine 17,860 stars that are located in the cluster vicinity up to a limit of $V=20$ mag. They showed that the colour excess in the direction of cluster field ranges over $0.46\leq E(B-V)\leq  0.80$ mag and calculated the mean distance of NGC 7654 as $d=1380\pm 70$ pc. Using the isochrones of \citet{Bertelli94}, \citet{Pandey01} calculated the age of cluster as $t=160$ Myr and determined the mass function slope as $X=-1.40\pm 0.07$ for the main-sequence stars in the mass range $0.8<M/M_{\odot}<4.5$. NGC 7654 was also studied using the {\it Gaia} DR2 data \citep{Gaia18}: \citet{Cantat-Gaudin18} gave the distance of NGC 7654 as $d=1600\pm263$ pc, and mean proper motion components as $\mu_{\alpha}\cos\delta=-1.94\pm0.01$ and $\mu_{\delta}=-1.13\pm0.01$ mas yr$^{-1}$;  \citet{Liu19} estimated distance, age and metal abundance of NGC 7654 as $d=1681\pm165$ pc, $t=380\pm23$ Myr and $\rm \log(Z/Z_{\odot})=-0.75$, respectively. In addition to this, they measured the mean proper motion components of the cluster as $\mu_{\alpha}\cos\delta=-1.93\pm0.01$ and $\mu_{\delta}=-1.13\pm0.01$ mas yr$^{-1}$. \citet{Cantat-Gaudin20} derived a distance estimate based on isochrone fitting as $d= 1653$ pc, and from the trigonometric parallaxes of member stars as $d=1678\pm130$ pc. The age and interstellar extinction of the cluster are given in the study as $t=155$ Myr and $A_{\rm V}=1.85$ mag (which corresponds to $E(B-V)=0.60$ mag) respectively (see Table 1).

\begin{figure}
\centering
\includegraphics[scale=.43, angle=0]{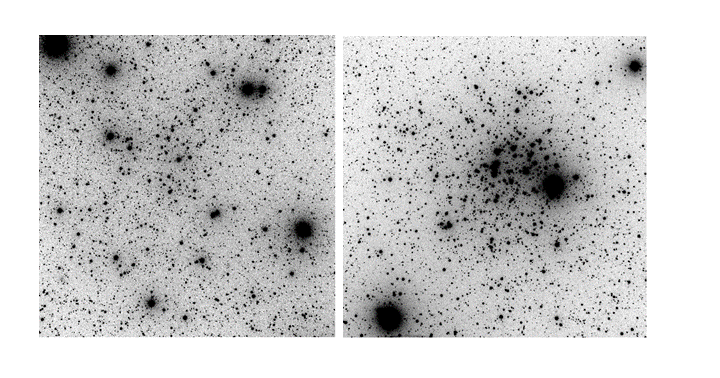}
\caption{Inverse coloured (negative) $21'\times 21'$ $V$-band images of the two clusters: the total exposure time is 1200 sec for Czernik 2 (left panel) and 800 sec for NGC 7510 (right panel). In these printed images, East and North correspond to the left and up directions, respectively.} 
\end {figure}

\begin{table*}[htbp]
  \centering
  \caption{Observation log:  columns denote name of clusters, observation date, filters, exposure times (in seconds), and the number of exposures ($N$). Dates are day-month-year.}
    \begin{tabular}{ccccc}
    \hline
    &       & \multicolumn{3}{c}{Filter/Exposure Time (s) $\times N$} \\
    \hline
    Cluster & Obs. Date & $U$ & $B$ & $V$ \\
    \hline
    Czernik 2 & 06.10.2018 & 90$\times$5, 900$\times$3 & 8$\times$5, 90$\times$5, 900$\times$2  & 5$\times$5, 60$\times$5, 600$\times$2 \\
    NGC 7654  & 07.10.2018 & 30$\times$1, 50$\times$5, 600$\times$3, 1200$\times$4 & 3$\times$5, 30$\times$5, 400$\times$4 & 1.5$\times$5, 15$\times$5, 200$\times$4 \\
    \hline
    \end{tabular}%
\end{table*}%

\section{Observations}
\label{sect:Obs}

Photometric observations of Czernik 2 and NGC 7654 were made using Bessell {\em UBV} filters with the 1-m T100 telescope located at T\"UB\.ITAK National Observatory (TUG) in Antalya, Turkey. The T100 is a  Ritchey Chretien telescope, which was equipped for this project with a Spectral Instruments 1100 Cryo-cooler cooled CCD camera placed in the focal plane. The CCD camera is a 4K$\times$4K pixel Fairchild model 486 BI (back-illuminated and with improved ultraviolet sensitivity). The physical scale of the pixels making up the CCD chip is 15$\times$15 microns. The pixel scale of the telescope-camera system is $0^{''}.31$ pixel$^{-1}$, leading to a $21'\times21'$ field of view on the sky. The CCD gain is 0.57 electrons ADU$^{-1}$ with a readout noise of 4.11 electrons. The ``bias'' level of CCD camera is around 500 ADU, the well capacity is about 142,900 electrons, the dark current is 0.0002 e$^-$ per pixel per second, and the CCD can be cooled down to $-100^{\rm o}$ C reducing the dark current to the reported level. The telescope-camera system is advantageous in that it has a wide field of view which allows observation of many open star clusters in a single exposure and has high quantum efficiency at both long and short wavelengths. During the observations we used different exposure times in {\em UBV} filters so as to obtain images of bright and faint stars in the regions of clusters with high signal to noise ($S/N$) ratio and to avoid saturation of CCD pixels. See Fig.~1 for example $V$-band images with inverted colours of the two clusters. See Table 2 for the observation log. Photometric calibrations were based on \citet{Landolt09} standard stars in a total of 14 areas whose observation details are given in Table 3. Frames were stacked. We utilised IRAF's standard CCD calibration processes, and applied PyRAF and astrometry.net routines for astrometric corrections to all cluster images. We computed instrumental magnitudes of standard stars via aperture photometry using the relevant IRAF packages. Then we applied multiple linear regression to these magnitudes to obtain photometric extinction and transformation coefficients for the two nights of observation (see Table 4). Instrumental magnitudes of the stars in cluster direction were measured with PSF Extractor (PSFEx) routines \citep{Bertin96}. Aperture corrections were then applied to these magnitudes before finally transforming instrumental magnitudes to standard brightnesses in the \citet{Johnson53} photometric system  via the transformation equations given by \citet{Janes11}. 

\begin{table}[htbp]
\setlength{\tabcolsep}{3pt}
  \centering
  \caption{Information on the observations of standard stars from selected \citet{Landolt09} fields. The columns are the observation date, star field name as from Landolt, the number of standard stars ($N_{\rm st}$) observed in a given field, the number of pointings to each field ($N_{\rm obs}$, i.e., observations), and the airmass range the fields (on a given night) were observed over ($X$). Dates are day-month-year.}

    \begin{tabular}{lcccc}
    \hline
Date	  & Star Field	& $N_{\rm st}$	& $N_{\rm obs}$	& $X$\\
\hline
          & SA92  & 6         & 2	        &          \\
	      & SA93	& 4	        & 1	        &          \\
	      & SA95	& 9 	    & 1	        &          \\
06.10.2018 & SA96	& 2	        & 1	        & 1.230 - 2.017\\
	      & SA98	& 19	    & 1	        &          \\
	      & SA100	& 5	        & 1	        &          \\
	      & SA114	& 5	        & 1	        &          \\
\hline	      
      	  & SA93	& 4	        & 1	        &          \\
      	  & SA94	& 2	        & 2	        &          \\
      	  & SA96	& 2	        & 1	        &          \\
      	  & SA97  & 2	        & 2	        &          \\
      	  & SA98	& 19	    & 1	        &          \\
07.10.2018 & SA99	& 3	        & 1	        & 1.242 - 1.872 \\
          & SA109	& 2	        & 1	        &          \\
	      & SA110	& 4	        & 2         &          \\	
	      & SA112	& 6	        & 2	        &          \\
	      & SA113	& 15	    & 1	        &          \\
          & SA114	& 5	        & 1	        &          \\

    \hline
    \end{tabular}%
\end{table}%

\begin{table*}[htbp]
  \centering
  \caption{Transformation and extinction coefficients derived for the two observation nights: $k$ and $k'$ are the primary and secondary extinction coefficients, respectively. $\alpha$ and $C$ are the transformation coefficients. Dates are day-month-year.}
    \begin{tabular}{lccccc}
    \hline
    Filter/Colour index & Obs. Date       & $k$     & $k'$    & $\alpha$     & $C$ \\
    \hline
$U$     & 06.10.2018 & 0.5795$\pm$0.0624 & -0.0747$\pm$0.0793 & ---  & --- \\
$B$     &           & 0.2620$\pm$0.0467 & -0.0303$\pm$0.0567 & 0.9775$\pm$0.0842 & 1.3908$\pm$0.0714 \\
$V$     &           & 0.1482$\pm$0.0189 & ---  & ---  & --- \\
$U-B$   &           & ---               & ---  & 0.9881$\pm$0.1196 & 3.5383$\pm$0.0968 \\
$B-V$   &           & ---               & ---  & 0.1009$\pm$0.0106 & 1.4750$\pm$0.0312 \\
$U$     & 07.10.2018 & 0.2154$\pm$0.0588 & +0.1071$\pm$0.0766 & ---  & --- \\
$B$     &           & 0.1651$\pm$0.0458 & +0.0616$\pm$0.0520 & 0.8214$\pm$0.0821 & 1.5979$\pm$0.0713 \\
$V$     &           & 0.1212$\pm$0.0164 & ---  & ---  & --- \\
$U-B$   &           & ---               & ---  & 0.7120$\pm$0.1161 & 4.1107$\pm$0.0883 \\
$B-V$   &           & ---               & ---  & 0.0773$\pm$0.0069 & 1.5709$\pm$0.0253 \\
\hline
    \end{tabular}%
\end{table*}%

\section{Data Analysis}
\subsection{Photometric Data}

The photometric catalogues of Czernik 2 and NGC 7654 list all identified stars located in the cluster areas, providing information such as positions ($\alpha$, $\delta$), apparent $V$ magnitude, $U-B$ and $B-V$ colours indices, proper motion components ($\mu_{\alpha}\cos\delta$, $\mu_{\delta}$), trigonometric parallaxes ($\varpi$) from {\it Gaia} EDR3, and membership probabilities ($P$). Photometric errors for Johnson ($V$, $U-B$, $B-V$) and {\it Gaia} EDR3 ($G$, $G_{\rm BP}-G_{\rm RP}$) magnitudes and colour indices are taken as internal errors. Table 5 lists the mean photometric errors in consecutive $V$-magnitude intervals. At $V=21$ mag the mean internal errors of photometric measurements for Czernik 2 reach 0.04 for the $V$-band magnitude and 0.07 mag for the $B-V$ colour index. For stars brighter than $V=20$ mag the errors reach 0.06 mag for the $U-B$ colour index. Similarly, for NGC 7654 these mean internal errors are about 0.04, 0.07, and 0.07 mag for $V$ magnitude, $U-B$, and $B-V$ colour indices, respectively (see Table 5). The mean photometric errors are smaller than 0.003 and 0.04 mag in $G$ -band and $G_{\rm BP}-G_{\rm RP}$ colour index, respectively, for the stars brighter than $V=20$ mag for Czernik 2. Similarly, these errors reach up to 0.003 mag in $G$-band and 0.20 mag in the $G_{\rm BP}-G_{\rm RP}$ colour index for the stars within $V=20$ mag in NGC 7654.

\begin{table*}[htbp]
\setlength{\tabcolsep}{4pt}
  \centering
  \caption{The mean internal photometric errors ($\sigma_{\rm V}$, $\sigma_{\rm U-B}$, $\sigma_{\rm B-V}$, $\sigma_{\rm G}$, $\sigma_{G_{\rm BP}-G_{\rm RP}}$) and number of measured stars ($N$) in the corresponding $V$ magnitude interval for each cluster.}
    \begin{tabular}{ccccccc|cccccc}
      \hline
    \multicolumn{7}{c}{Czernik 2} & \multicolumn{6}{c}{NGC 7654} \\
    \hline
  $V$ & $N$ & $\sigma_{\rm V}$ & $\sigma_{\rm U-B}$ & $\sigma_{\rm B-V}$ & $\sigma_{\rm G}$ &  $\sigma_{G_{\rm BP}-G_{\rm RP}}$ & $N$ & $\sigma_{\rm V}$ & $\sigma_{\rm U-B}$ & $\sigma_{\rm B-V}$ & $\sigma_{\rm G}$ & $\sigma_{G_{\rm BP}-G_{\rm RP}}$\\
  \hline
  ( 7, 12] & 14  & 0.001 & 0.002 & 0.001 & 0.004 & 0.006 &   20 & 0.001 & 0.001 & 0.001 & 0.003 & 0.005\\
  (12, 14] & 58  & 0.002 & 0.005 & 0.002 & 0.003 & 0.005 &  101 & 0.001 & 0.002 & 0.001 & 0.003 & 0.009\\
  (14, 15] & 111 & 0.002 & 0.006 & 0.002 & 0.003 & 0.005 &  131 & 0.001 & 0.003 & 0.002 & 0.003 & 0.010\\
  (15, 16] & 163 & 0.001 & 0.008 & 0.002 & 0.003 & 0.006 &  184 & 0.002 & 0.007 & 0.003 & 0.003 & 0.011\\
  (16, 17] & 313 & 0.002 & 0.014 & 0.004 & 0.003 & 0.006 &  366 & 0.003 & 0.013 & 0.005 & 0.003 & 0.011\\
  (17, 18] & 587 & 0.004 & 0.028 & 0.007 & 0.003 & 0.010 &  584 & 0.006 & 0.025 & 0.009 & 0.003 & 0.186\\
  (18, 19] & 965 & 0.008 & 0.046 & 0.014 & 0.003 & 0.020 & 1023 & 0.012 & 0.047 & 0.020 & 0.003 & 0.219\\
  (19, 20] & 1435& 0.018 & 0.056 & 0.029 & 0.003 & 0.040 & 1262 & 0.026 & 0.070 & 0.044 & 0.003 & 0.201\\
  (20, 21] & 1431& 0.041 & ---   & 0.070 & 0.004 & 0.082 &  176 & 0.043 & 0.073 & 0.072 & 0.004 & 0.060\\
      \hline
    \end{tabular}%
\end{table*}%

\begin{figure*}
\centering
\includegraphics[scale=0.55, angle=0]{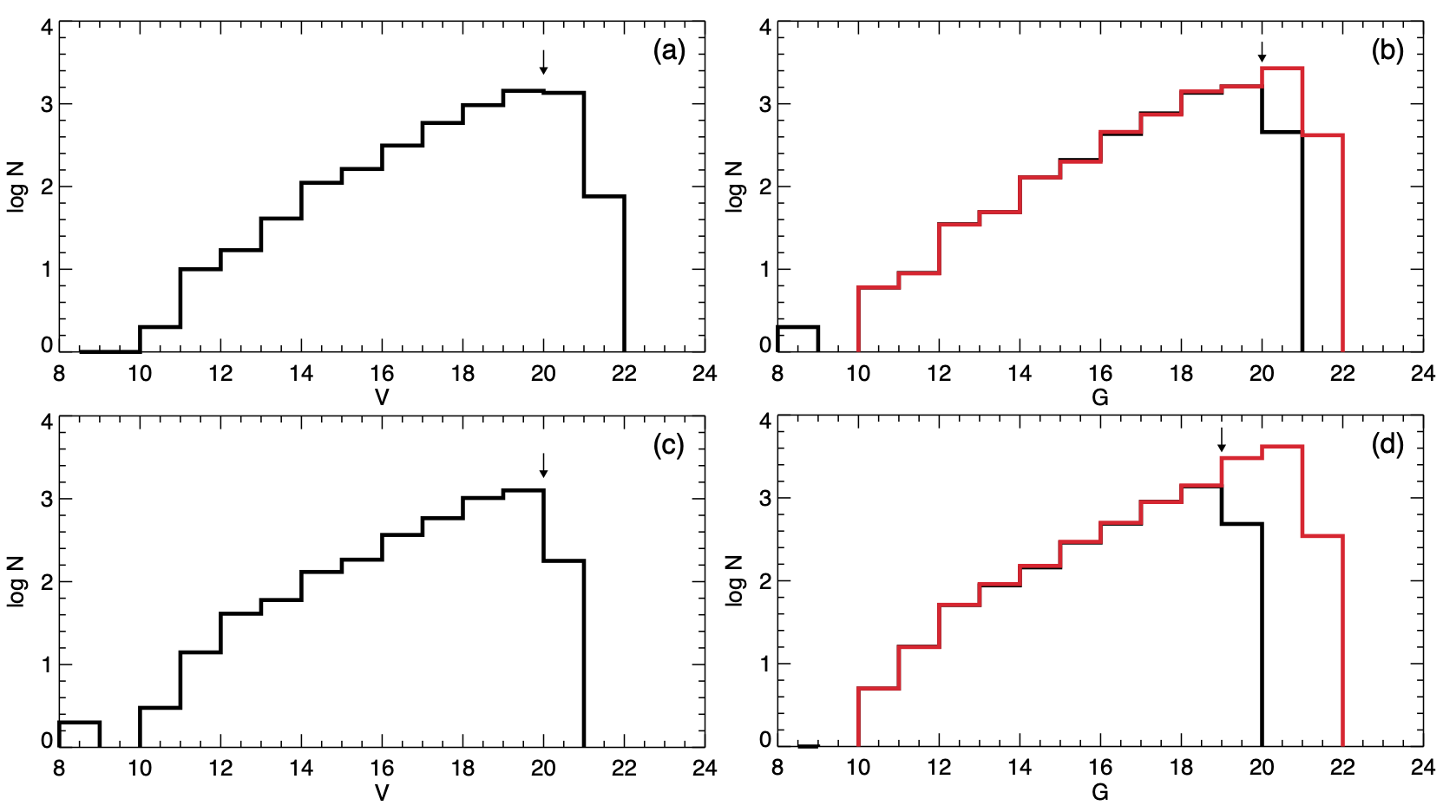}\\
\caption{Interval $V$ and $G$-band magnitude histograms of Czernik 2 (a, b) and NGC 7654 (c, d): The arrows show the faint limiting apparent magnitudes in $V$ and $G$-bands. Black lines indicate the star counts based on the stars detected in the study, while red lines are star counts based on the stars taken from {\it Gaia} EDR3 for the same cluster regions.} 
\end {figure*} 

\begin{figure*}
\centering
\includegraphics[scale=0.36, angle=0]{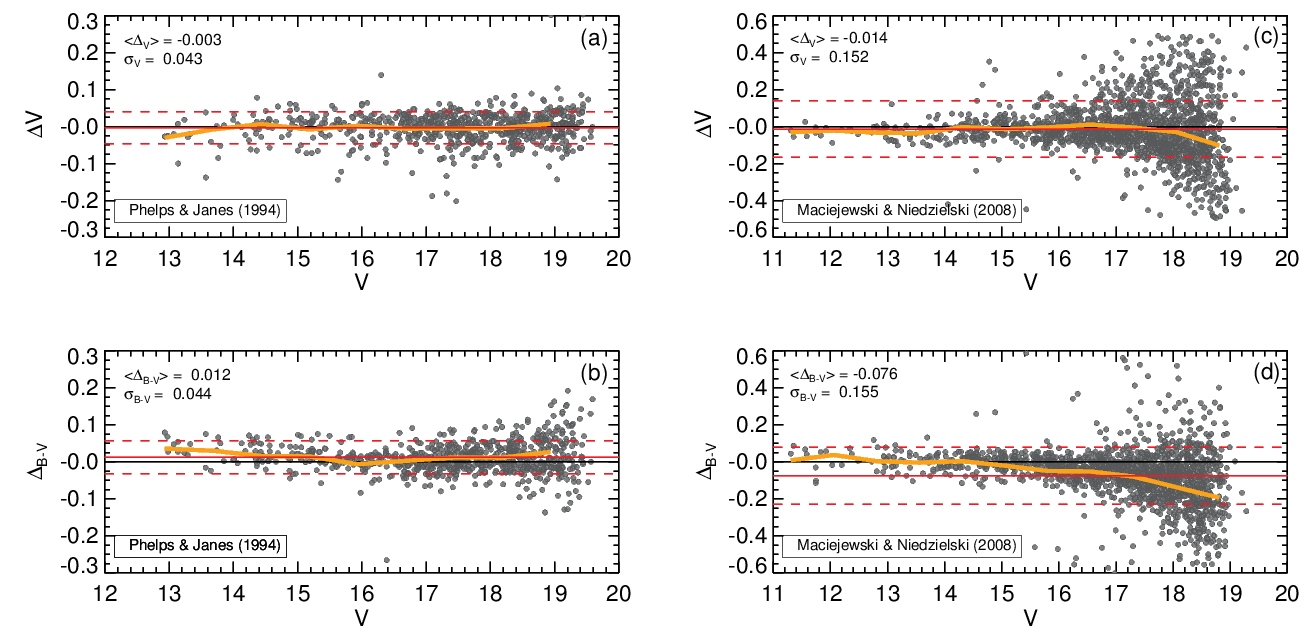}\\
\caption{Comparisons of observational magnitude and colours with those calculated from \citet{Phelps94} (a-b) and \citet{Maciejewski08} (c-d) for Czernik 2. The mean differences and their $\pm 1\sigma$ dispersions are represented with solid and dashed red lines, respectively. Yellow solid lines indicate the trends of moving average of data}.
\end {figure*} 

The exposure times used during the observations contribute toward (together with crowding) the detection limit for a star in the observed star field. In order to obtain this completeness limit, we constructed $V$ and $G$ magnitude histograms for Czernik 2 and NGC 7654 in which the number of stars is a function of $V$  and $G$ magnitude intervals (see Fig.~2). To check photometric completeness limits we took the stars for the region of each cluster using {\it Gaia} EDR3 data. Taking into account central equatorial coordinates \citep{Cantat-Gaudin20} and using the same area as the CCD images ($21'\times 21'$) of two clusters, we identified the stars within $8 < V < 24$ mag range for each cluster region. Using these stars we constructed histograms of $G$-magnitude intervals, displayed as Figures 2b and 2d.  In Figure 2 the solid black lines denote the  observational values, while the red lines (Figures 2b and 2d) represent the stars from {\it Gaia} EDR3.  We can clearly see from Figures 2b and 2d that the two distributions are in good agreement before the completeness limit is met for our observational data. Considering the turnover points from which the number of detected stars drops with increasing magnitude, we adopted completeness limits as being $V=20$ mag for both clusters. This limit corresponds to $G=20$ mag for Czernik 2 and $G=19$ mag for NGC 7654 where the number of {\it Gaia} EDR3 stars starts to overrun the number of observed ones. The use of different telescope-detector combinations in observations affects the detected number of stars due to changes in telescope detection limits, particularly at fainter magnitudes. This could result in detecting more stars at fainter magnitudes ($G>20$) with {\it Gaia} observations in the two cluster areas (Figs.\ 2b \& 2d). It can be seen in Fig.~2 that incompleteness (of stellar detections) becomes significant beyond the completeness limits. Therefore further analyses did not consider stars fainter than $V=20$ mag for both Czernik 2 and NGC 7654.

We compared our photometry with those from previous studies employing the same filter system. We reviewed CCD {\em UBVI} data for NGC 7654 given by \citet{Pandey01} and CCD {\em BV} data for Czernik 2 presented in two studies \citep{Phelps94, Maciejewski08}. 724 and 1,589 of the 5,077 stars detected in the direction of Czernik 2 were matched (via equatorial coordinates) with the stars in catalogues of \citet{Phelps94} and \citet{Maciejewski08}, respectively. We plotted apparent $V$ magnitude versus differences between our $V$ and $(B-V)$ measurements and those given in the two studies (see Fig. 3). As seen in the figure, there is no bias with \citet{Phelps94} and up to $V=17$ mag with \citet{Maciejewski08}. The mean difference and standard deviation of $V$ magnitudes and the $(B-V)$ colour index are derived as $\langle \Delta V\rangle=-0.003$ mag, $\langle \sigma_{\rm V}\rangle=0.043$ mag, $\langle \Delta (B-V)\rangle = 0.012$ mag, and $\sigma_{\rm B-V}=0.044$ mag for the comparison with \citet{Phelps94}. Comparison with \citet{Maciejewski08}'s data led to $\langle \Delta V\rangle=-0.014$ mag, $\langle \sigma_{\rm V}\rangle=0.152$ mag, and $\langle \Delta (B-V)\rangle = -0.076$ mag, $\sigma_{\rm B-V}=0.155$ mag. We also calculated a moving average from the comparison with each data to understand the trend of mean differences, presented as yellow solid lines in Fig. 3. For the comparison with \citet{Phelps94}, the moving average trends are near zero in $V$ magnitude and $B-V$ colour index (Fig. 3a, b). This trend for the comparison of $V$ magnitudes is nearly zero up to $V=18$ mag (Fig. 3c). Similarly, the moving average trend in the $(B-V)$ colour index is nearly zero up to $V=15.5$ mag, while it goes towards nearly 0.2 mag at the fainter magnitudes (Fig. 3d). Consequently, the mean differences in apparent magnitude and colour indices, together with their small standard deviations, show that our photometric data are well-matched with those of \citet{Phelps94}.

We cross-matched our catalogue with the catalogue of \citet{Pandey01} to compare the photometric measurements. 3,365 of the equatorial coordinate data of 3,847 stars detected in our NGC 7654 area were matched with the stars in the catalogue presented by \citet{Pandey01}. Plots of apparent $V$ magnitude versus differences between our $V$, $(U-B)$, $(B-V)$ measurements and those given by \citet{Pandey01} are shown in Fig. 4. There is no bias up to $V=18$ mag with the study of \citet{Pandey01}. A small trend can be seen for the magnitude fainter than $V= 18$. Based on this comparison the mean difference and standard deviation in $V$ are derived as $\langle \Delta V\rangle=0.046$ mag and $\langle \sigma_{\rm V}\rangle=0.148$ mag, respectively. While the mean differences up to $V=13.5$ mag in the $(B-V)$ colour index are about zero, these differences increase towards approximately 1.0 mag at the fainter magnitudes (Fig. 4b). The mean difference and standard deviation of the $(B-V)$ colour index are derived as $\langle \Delta (B-V)\rangle = 0.059$ mag and $\sigma_{\rm B-V}=0.215$ mag, respectively. Fig. 4c gives a comparison of the $(U-B)$ colour index. The mean difference and standard deviation of the comparison of the $(U-B)$ colour index are derived as $\langle \Delta(U-B)\rangle=0.101$ mag and $\sigma_{\rm U-B}=0.269$ mag, respectively. Moving average trends are about zero for each comparison. Although the mean errors of the $(U-B)$ colour index are small, their standard deviations increase towards the fainter magnitudes as expected. Overall, our photometric data are in good agreement with the study of \citet{Pandey01}.
\begin{figure}[htbp]
\centering
\includegraphics[scale=0.35, angle=0]{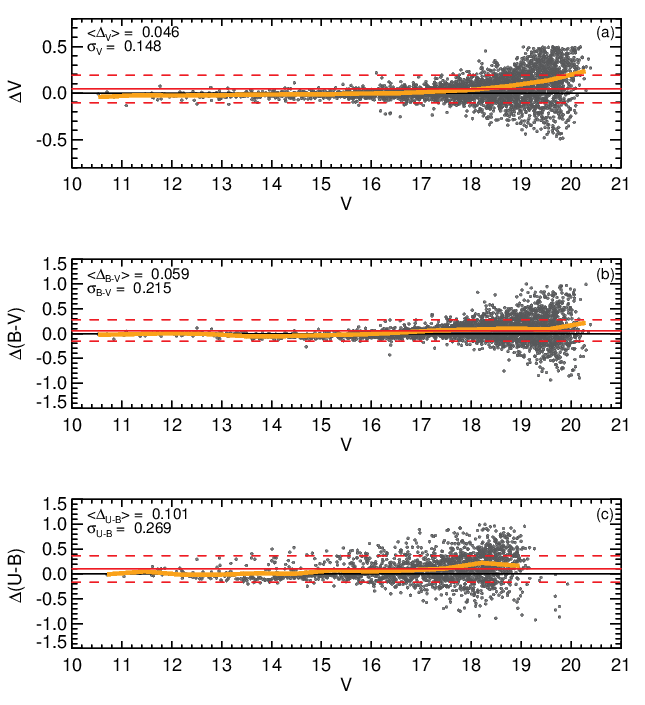}\\
\caption{Comparison of photometric measurements in this study with data of \citet{Pandey01} (a-c) as a function of $V$ magnitudes for NGC 7654. The mean differences and their $\pm 1\sigma$ dispersions are represented with solid and dashed red lines, respectively. Yellow solid lines indicate the trends of moving average of data.} 
\end {figure} 

\begin{figure}[htbp]
\centering
\includegraphics[scale=0.43, angle=0]{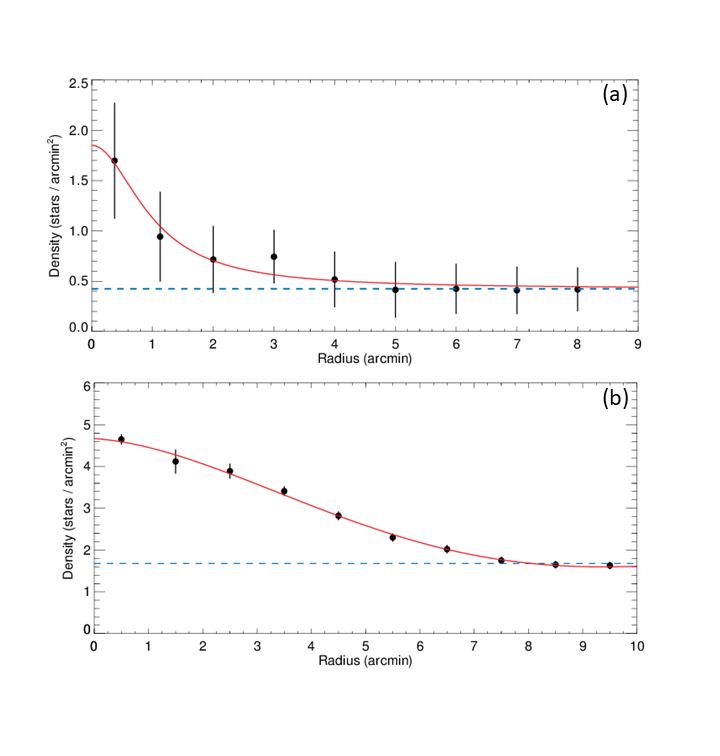}\\
\caption{The radial density profiles for the Czernik 2 (a) and NGC 7654 (b) clusters. Errors were derived using equation of $1/\sqrt N$, where $N$ represents the number of stars used in the density estimation. King profile and background stellar density are presented with solid red curve and blue dashed line, respectively.} 
\end {figure} 

\subsection{Spatial Structure of the Clusters}
The morphologies of Czernik 2 and NGC 7654 are different to each other: NGC 7654 shows a spherical structure with a heavy central concentration, whereas the irregular shape and sparse stellar distribution of Czernik 2 make it difficult to separate from the surrounding star field. Taking central coordinates from the study of \citet{Cantat-Gaudin20} and utilising the radial density profile (RDP) as defined by \citet{King62}, we determined the structural parameters of two clusters. We divided a given cluster's area into a series of concentric circles of one arcmin centred on the cluster centre. Additionally we estimated the stellar density ($\rho$) by taking the ratio of the number of stars in a given annulus to the area of that annulus. For each cluster we plotted density versus radius, fitting the RDP \citep{King62} using a $\chi^2$ minimisation method. \citet{King62} described the central density $\rho(r)$ of a cluster as $\rho(r)=f_{bg}+[f_0/(1+(r/r_c)^2)]$ where $r$ and $r_c$ are a given radius from the cluster centre and core radius respectively, $f_0$ and $f_{bg}$ are the central and background densities. The best fits to the RDPs of Czernik 2 and NGC 7654 are shown in Fig. 5. The solid red curves denote the King profiles and the blue dashed lines represent the background densities. The levels of the modeled number density profiles equals the background densities  $r=5$ arcmin from the centre of Czernik 2 and $r=8$ arcmin from the centre of NGC 7654 (see Fig. 5). We considered the stars within these radius values for further analysis. These adopted radius levels for the two clusters are in good agreement with the study of \Citet{Cantat-Gaudin18}, who presented the clusters' radii ($r_{50}$) which contain half of the stars in the clusters.  Applying the formulae of \citet{King62}, we found the central stellar densities to be $f_0=1.426~\pm 0.663$ and $f_0=3.024~\pm 0.274$ stars arcmin$^{-2}$, background stellar densities to be $f_{bg}=0.424~\pm 0.266$ and $f_{bg}=1.680~\pm 0.257$ stars arcmin$^{-2}$ and core radii to be $r_c=0.988\pm 0.879$ and $r_c=4.663\pm0.699$ arcmin for Czernik 2 and NGC 7654, respectively (Table 6).

\begin{table}[htbp]
\setlength{\tabcolsep}{1.6pt}
\small{
  \centering
  \caption{Results of RDP fitting for the two clusters: $f_0$, $f_{bg}$ and $r_c$ represent central stellar density, background stellar densities and the core radius, respectively.}
    \begin{tabular}{cccc}
    \hline
    Cluster & $f_0$  & $f_{bg}$  & $r_c$ \\
            & (stars arcmin$^{-2}$) & (stars arcmin$^{-2}$) & (arcmin) \\
    \hline
    Czernik 2 & 1.426$\pm$0.663 & 0.424$\pm$0.266 & 0.988$\pm$0.879  \\
    NGC 7654  & 3.024$\pm$0.274 & 1.680$\pm$0.257 & 4.663$\pm$0.699  \\
    \hline
    \end{tabular}%
    }
\end{table}%

\subsection{CMDs and Membership Probabilities of Stars}
It is important in cluster studies to estimate the field star contamination. In order to get reliable parameters for an open cluster, it is necessary to identify physical member stars of it accurately. Sharing the same origin allows member stars of the cluster to have co-movements together in the sky, concentrated in the proper motion space unlike the neighbouring field stars. These features make it possible to separate cluster members from field star contamination using their proper motion values \citep{Angelo19, Angelo20}.

\begin{figure*}
\centering
\includegraphics[scale=0.27, angle=0]{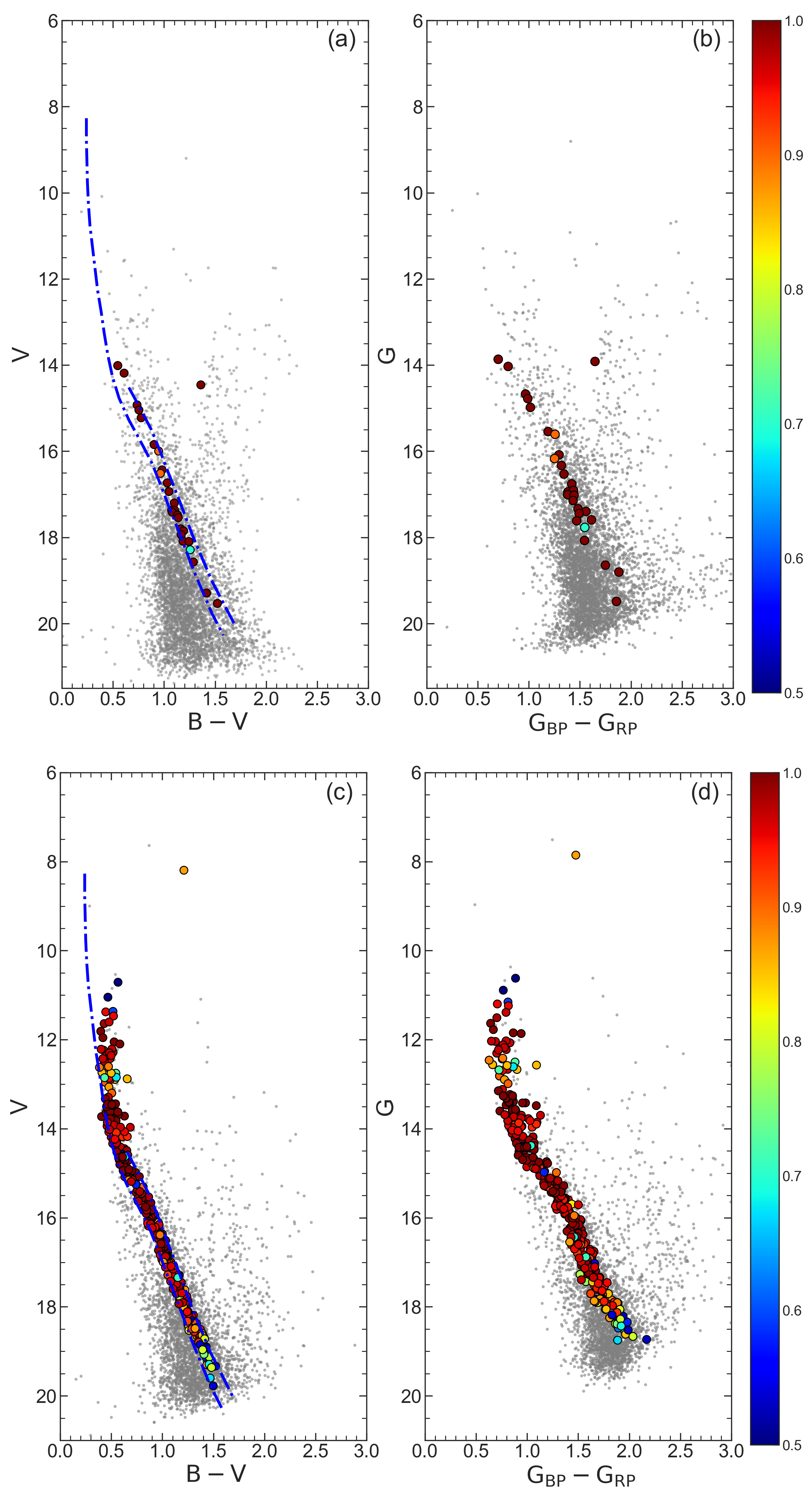}
\caption{$V\times (B-V)$ and $G\times (G_{\rm BP}-G_{\rm RP})$ CMDs of Czernik 2 (a, b) and NGC 7654 (c, d). The blue dot-dashed lines represent the ZAMS \citep{Sung13} including binary star effect. The membership probabilities of stars that lie within the fitted ZAMS are shown with different colours, these member stars are located within $r\leq 5'$ and $r\leq 8'$ of the cluster centres calculated for Czernik 2 and NGC 7654, respectively. Grey dots indicate field stars.} 
\end {figure*}

In this study, we used {\it Gaia} EDR3 astrometric data \citep{Gaia20} and utilised the UPMASK \citep[Unsupervised Photometric Membership Assignment in Stellar Clusters; ][]{Krone-Martins14} method to calculate cluster membership probabilities of the observed stars. UPMASK describes an open star cluster as a group of stars with a common origin taking into account their astrometric data (proper motion components and trigonometric parallaxes) and defines the field stars as spatially `scattered' objects with different origins. This assumption by UPMASK is key for the method to determine the membership probabilities of stars and to separate cluster members from field stars. The method uses a machine-learning algorithm, $k$-means clustering, to find spatially concentrated groups and identify member stars of the cluster. Here, $k$ is the number of clusters and is not adjusted directly by the user. The number of the clusters is identified by dividing the total number of stars by the number of stars per cluster, converting the result to an integer in each iteration \citep{Krone-Martins14}. As mentioned by \citet{Krone-Martins14} and \citet{Cantat-Gaudin18}, the best result from this technique is achieved when the integer value is within 10 to 25. Recently, \citet{Cantat-Gaudin_Anders20}  used {\it Gaia} DR2 astrometric data together with the UPMASK methodology on 1,481 open clusters,  successfully calculating stellar membership probabilities in these clusters as a continuation of the study of \citet{Cantat-Gaudin18}. In the studies of \citet{Cantat-Gaudin18} and \citet{Cantat-Gaudin_Anders20}, the mean $k$-means values across the clusters were both 10.

In this study, we utilised the UPMASK method according to a five-dimensional astrometric space ($\alpha$, $\delta$, $\varpi$, $\mu_{\alpha}\cos\delta$, $\mu_{\delta}$) and applied it to calculate the membership probabilities ($P$) of stars for the two clusters. We scaled these five observable measures to unit variance and ran for each cluster 100 iterations assessing cluster membership. The membership probability is defined as the frequency with which a star is marked as a part of a clustered group. Best results were reached when $k$ was set to 15 for both clusters. Stars with membership probabilities $P\geq 0.5$ were chosen as the most likely member of clusters. We found 154 and 530 potential member stars for Czernik 2 and NGC 7654, respectively, for subsequent analysis. Examining the two open clusters, \citet{Cantat-Gaudin18} and \citet{Liu19} give the number of stars with cluster membership probabilities $P\geq 0.5$, as 102 and 95 for Czernik 2, and 1,063 and 1,061 for NGC 7654 respectively.  Considering both studies, \citet{Cantat-Gaudin18} used the UPMASK method with {\it Gaia} DR2 data, while \citet{Liu19} applied the SHiP method considering {\it Gaia} DR2 astrometric data in the calculation of the cluster membership probabilities of the stars. The reason of the differences between these studies and the current one is that in this study we used current {\it Gaia} EDR3 astrometric data during membership determinations, while \citet{Cantat-Gaudin18} and \citet{Liu19} took into account the astrometric data of {\it Gaia} DR2. It is worth noting that while UPMASK determines the membership probabilities of stars, it assumes that stars are single and therefore cannot be used to identify probable binary stars. For this reason, beside the membership probability criteria, membership selection additional information was taken into account to account for binaries. We plotted the $V\times (B-V)$ CMDs for two clusters and fitted the zero age main-sequence \citep[ZAMS, ][]{Sung13} to these diagrams. The lower base of the cluster main-sequence was determined by fitting the ZAMS by eye according to the most probable member stars ($P\geq 0.5$), then to make it possible to select the most probable binary stars we shifted the fitted ZAMS 0.75 mag to the brighter magnitudes (Fig. 6). To determine basic parameters of the two clusters, we considered the most probable member stars to be those that lie between the fitted ZAMS lines and are located within the cluster limiting (outer) radii estimated in the study (mentioned in Section 3.2). Thus, we arrived at 28 and 369 ``member'' stars for Czernik 2 and NGC 7654, respectively. Probability histograms for the stars (by cluster) are shown in Fig. 7. Vector-point diagrams (VPDs) were prepared to visualise the distributions of  most probable member stars in each cluster (see Fig. 8). Additionally, we estimated mean proper motion components of selected member stars, showing the intersection of these values on Fig. 8 with dashed lines.

\begin{figure*}
\centering
\includegraphics[scale=0.65, angle=0]{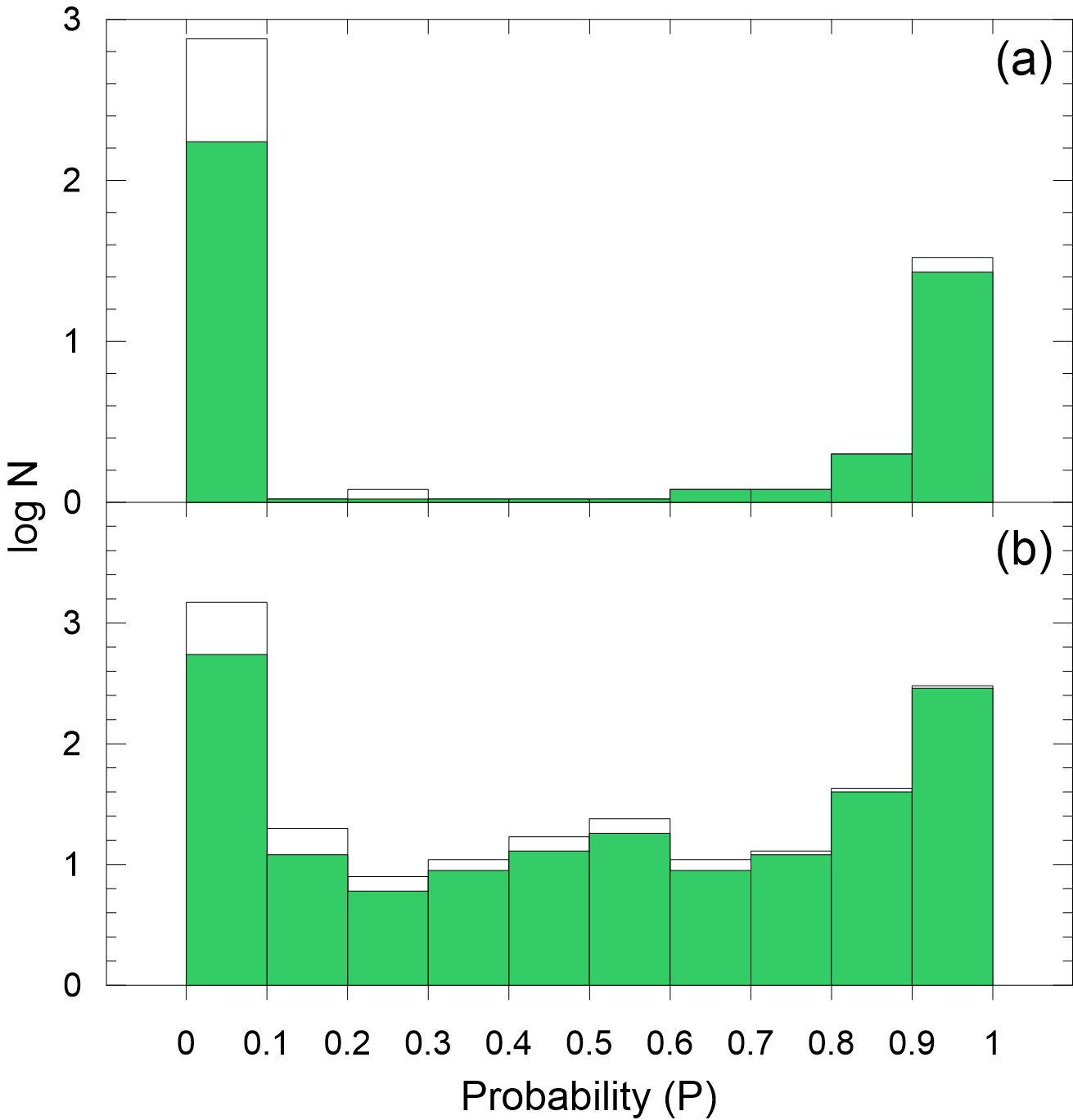}
\caption{The histogram of membership probabilities versus number of stars for Czernik 2 (a) and NGC 7654 (b). The green coloured shading denotes the stars that lie within the main-sequence band and effective cluster radii.} 
\end {figure*}

\begin{figure*}
\centering
\includegraphics[scale=0.34
, angle=0]{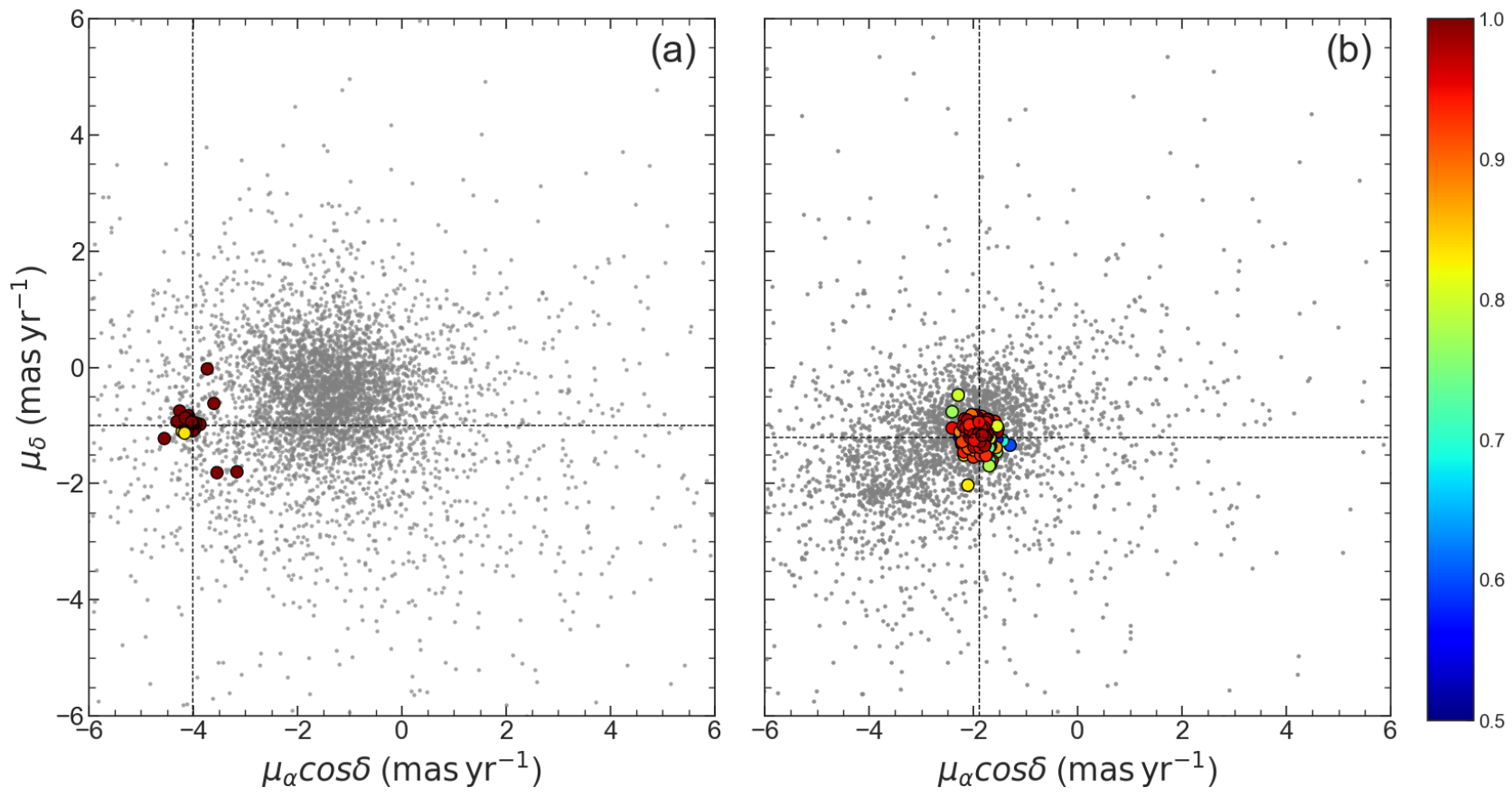}\\
\caption{Vector points diagrams of Czernik 2 (a) and NGC 7654 (b). Coloured dots identify the membership probabilities of the cluster stars according to colour scale shown on the right. Dashed lines are the intersection of the mean proper motion values. The colour scale shows the membership probabilities of the most likely cluster members.} 
\end {figure*}

\begin{figure*}
\centering
\includegraphics[scale=0.6, angle=0]{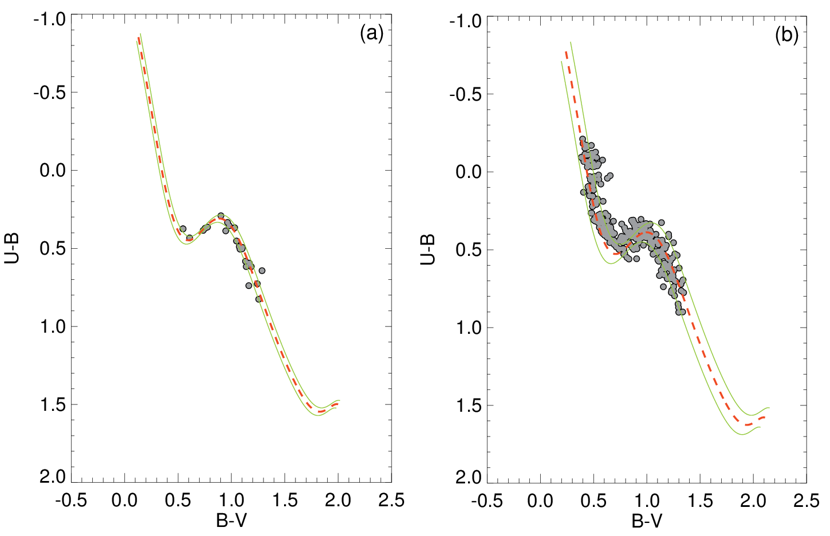}
\caption{$(U-B)\times (B-V)$ two-colour diagrams of the most probable member main-sequence stars in the regions of Czernik 2 (a) and NGC 7654 (b). Red dashed and green solid curves represent the reddened ZAMS given by \citet{Sung13} and $\pm1\sigma$ standard deviations, respectively.} 
\end {figure*}

\section{Astrophysical Parameters of the Clusters}

In this section we summarised the techniques applied during the cluster analyses (for detailed analyses on methods used see \citet{Bilir06, Bilir10, Bilir16, Yontan15, Yontan19, Yontan21,  Bostanci15, Bostanci18, Ak16, Banks20}). We estimated reddening and photometric metallicities from TCDs on an individual basis, then keeping these parameters as a constants we fitted theoretical isochrones on CMDs to determine distance moduli and age simultaneously. 

\subsection{Reddening}

In order to estimate reddening through the regions of two clusters, we selected the most probable main-sequence members whose $V$ magnitude ranges are within $12.5\leq  V\leq  18.5$ and $11.5\leq V\leq 19$ mag for Czernik 2 and NGC 7654, respectively. Using these sample stars, we constructed $(U-B) \times (B-V)$ diagrams and fitted an intrinsic ZAMS \citep{Sung13} on the observational data. The fitting procedure made steps of 0.001 mag utilising the $\chi^2$ minimisation technique until improvement was within a predefined limit. During the fitting process we adopted the slope of the reddening $E(U-B)/E(B-V)$ as 0.72 \citep{Johnson53} for both clusters. This procedure provided mean values of reddening as $E(B-V)=0.46\pm 0.02$ and $E(B-V)=0.57\pm 0.04$ mag for Czernik 2 and NGC 7654, respectively. Errors for the reddening are $\pm 1\sigma$ deviations. Best fit results with TCDs are shown in Fig. 9.

\citet{Cantat-Gaudin20} calculated the $V$ band absorptions of Czernik 2 and NGC 7654 open clusters as 1.41 and 1.85 mag, respectively. With the ratio of the $V$ band absorptions to the value of 3.1 belonging to the normal interstellar medium, the $E(B-V)$ colour excess for Czernik 2 and NGC 7654 were determined to be 0.45 and 0.60 mag, respectively. Considering the colour excess values of 0.46 and 0.57 mag calculated for Czernik 2 and NGC 7654 in this study, it seems that our results are consistent with those of \citet{Cantat-Gaudin20}.

\begin{figure*}
\centering
\includegraphics[scale=0.55, angle=0]{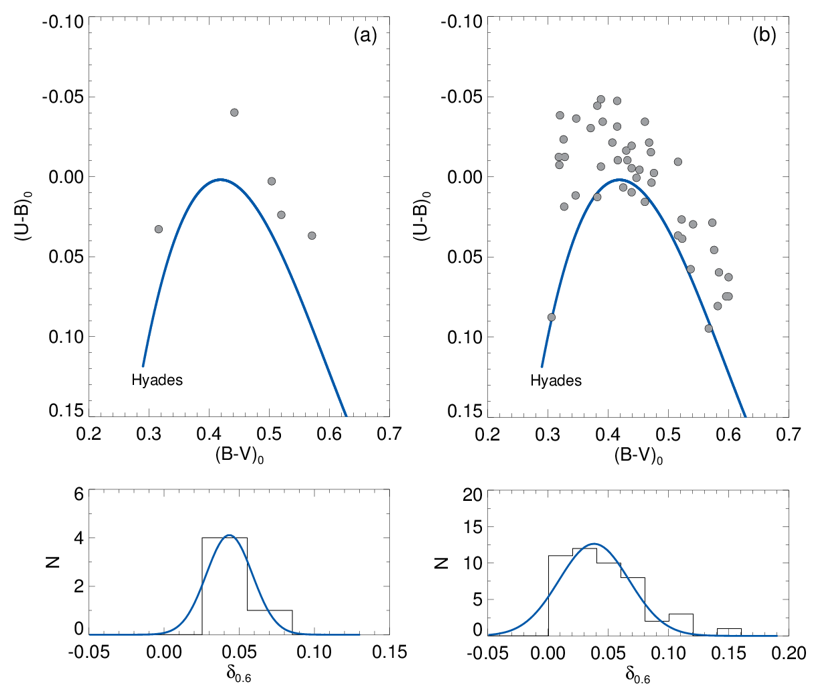}
\caption{TCDs (upper panel) and the distributions of normalised $\delta_{0.6}$ (lower panel) for Czernik 2 (a) and NGC 7654 (b). The solid blue lines in the upper and lower panels represent the main-sequence of Hyades and Gaussian models which were fitted to the histograms, respectively.} 
\end {figure*}

\subsection{Estimation of Metallicities}
We estimated photometric metallicities for the two clusters using their observational CCD {\em UBV} photometric data. The metallicities of Czernik 2 and NGC 7654 have not been estimated before. Applying the calibration given by \citet{Karaali11} on de-reddened TCDs of the two clusters, the metallicities were derived using the normalised UV excesses ($\delta (U-B)_{0.6}$) of the most probable member ($P\geq 0.5$) F-G type main-sequence stars within the colour index of $0.3\leq (B-V)_0\leq 0.6$ mag \citep{Eker18, Eker20}. $(U-B)_0\times(B-V)_0$ TCDs were plotted using F-G type cluster member stars and compared to the main-sequence of Hyades cluster. $(U-B)_0$ differences between the cluster and Hyades stars corresponding the same $(U-B)_0$ colour indices were calculated. These difference values are defined as UV excesses for each star, which can be written as $\delta =(U-B)_{\rm 0,H}-(U-B)_{\rm 0,S}$ \citep[$H$ and $S$ are the Hyades and cluster star, respectively, corresponding to an $(B-V)_0$ colour index, see also][]{Karaali03}. We normalised each UV excess ($\delta$) based on $(B-V)_0=0.6$ mag (i.e. $\delta_{\rm 0.6}$) and plotted the histogram of $\delta_{0.6}$ values. The mean $\delta_{0.6}$ was calculated by fitting a Gaussian to the distribution. We considered the mean $\delta_{0.6}$ value which represents the peak of optimal, fitted Gaussian. Based on $\delta_{0.6}$, we calculated the metallicities of each cluster using the equation improved by \citet{Karaali11} which is given as follow:
\begin{equation}
{\rm [Fe/H]}=0.105-3.557\times\delta_{0.6}-14.316\times\delta_{0.6}^2.
\end{equation} 
TCDs and histograms of normalised $\delta_{0.6}$ are shown in Fig. 10. The statistical uncertainty is the $\pm 1\sigma$ width of the optimal Gaussian model. We derived the metallicity for Czernik 2 as $\rm [Fe/H]=-0.08\pm 0.02$ dex from five member stars and for NGC 7654 as $\rm [Fe/H]=-0.05\pm 0.01$ dex from 47 member stars. 

The fractional abundance by mass, known as $Z$, is required to determine the age of a cluster. Bovy\footnote{ https://github.com/jobovy/isodist/blob/master/isodist/ Isochrone.py} analytically denoted an equation which transforms the [Fe/H] metallicities to the mass fraction $Z$, using PARSEC\footnote{PAdova and TRieste Stellar Evolution Code} models:
\begin{equation}
Z_{\rm X}={10^{{\rm [Fe/H]}+\log \left(\frac{Z_{\odot}}{1-0.248-2.78\times Z_{\odot}}\right)}},
\end{equation}      

\begin{equation}
Z=\frac{(Z_X-0.2485\times Z_{\rm X})}{(2.78\times Z_{\rm X}+1)}.
\end{equation} 
$Z_{\rm X}$ and $Z$ represent the intermediate operation function and all elements heavier than helium, respectively. $Z_{\odot}$ denotes the solar mass fraction, which we adopted as  0.0152 \citep{Bressan12}. Using the equations given above, we measured $Z=0.013$ and $Z=0.014$ for Czernik 2 and NGC 7654, respectively.

\subsection{Distance Moduli and Age Estimation}

We derived distance moduli and age estimation (for each cluster) simultaneously. We used classical isochro-nes fitting methods \citep[e.g.][]{Phelps94, Sharma06, Sharma17}. We fitted PARSEC isochrones \citep{Bressan12} to the observational $V\times (U-B)$,  $V\times(B-V)$, and $G\times (G_{\rm BP}-G_{\rm RP}$) CMDs consisting of the most probable member stars with $P\geq 0.5$. 

In the CMD morphology of the two clusters, the turn-off and main-sequence stars are significant, in addition, NGC 7654 has a star (BD +60 2532) in an advanced evolutionary stage. This star is important in estimating the age of NGC 7654. For this reason, it is necessary to examine the cluster membership  status of BD +60 2532 by considering its different parameters. According to astrometric data of BD+60 2532 in the {\it Gaia} EDR3 catalogue, its membership probability was determined to be $P=0.9$ and its distance to the cluster center as $r=3.81$ arcmin. In this study, the effective radius of the cluster is also determined to be $r=8$ arcmin. Moreover, \citet{Luck14} gives the atmosphere model parameters of the BD+60 2532 as $T_{\rm eff}=6114$ K, $\log g=3.75$  (cgs) and [Fe/H] = $-0.06$ dex. In our study we obtained the metallicity of the NGC 7654 as [Fe/H] = $-0.05 \pm 0.01$ dex, which is very well compatible with the result of \citet{Luck14}. These results show that the BD+60 2532 is most probably a member star of the cluster.

PARSEC models \citep{Bressan12} were selected with reference to the mass fractions ($Z$) measured for each cluster. The fitting procedure was done by eye keeping reddening and metallicity as constants. To obtain parameters carefully, we fitted PARSEC models considering main-sequence, turn-off and giant likely member stars in the cluster CMDs. To determine the errors in cluster ages, we fitted two more isochrones on CMDs at closer values (one younger and the other older than the best fit age) to the adopted cluster ages by considering the distribution of likely member stars in CMDs. Besides the best fit isochrones which give the expected ages of the clusters, these other closing fitting isochones provided insight into the likely errors in estimated ages. In addition to this, we calculated errors in distance moduli and distances using the relations presented in \citet{Carraro17}. When utilising the isochrone fitting procedure, we used reddening values, $E(B-V)$, determined earlier (see Sec. 4.1) for $V\times (U-B)$ and $V\times(B-V)$ CMDs. While estimating ages for the $G\times (G_{\rm BP}-G_{\rm RP})$ CMDs, the PARSEC isochrones were reddened considering the equation of $E(G_{\rm BP}-G_{\rm RP})=1.2803\times E(B-V)$ and extinction of $A_{\rm G}/A_{\rm V}=0.789$ whose the coefficients were inferred from the equations presented by \citet{Wang19}. The best fit $Z=0.013$ isochrone gives an age $t=1.2\pm 0.2$ Gyr and a distance modulus $\mu =12.80\pm 0.07$ mag for Czernik 2. The best fit $Z=0.014$ isochrone gives $t=120\pm 20$ Myr and a distance modulus $\mu =13.20\pm 0.16$ mag for NGC 7654. The distance modulus values correspond to the cluster distances being $d=1883\pm 63$ pc and $d=1935\pm 146$ pc for Czernik 2 and NGC 7654, respectively (See Table 1). Although 37 Myr represents the NGC 7654's main-sequence stars well, it seems that this model does not match the stars at the turn-off point. When BD+60 2532 is excluded from the statistics and the 120 Myr isochrone fitted according to cluster's most likely member stars, it is seen that this age better represents the turn-off and main sequence stars of the NGC 7654. In this case, it has been concluded that the position of the BD+60 2532 on CMDs may be due to the binary star effect and the radial velocity measurements of the star should be examined to elucidate this. Consequently, in this study we adopted the age of the NGC 7654 as $t=120\pm20$ Myr. The CMDs of two clusters with best fit isochrones are shown in Fig. 11.

\begin{figure*}
\centering
\includegraphics[scale=0.17, angle=0]{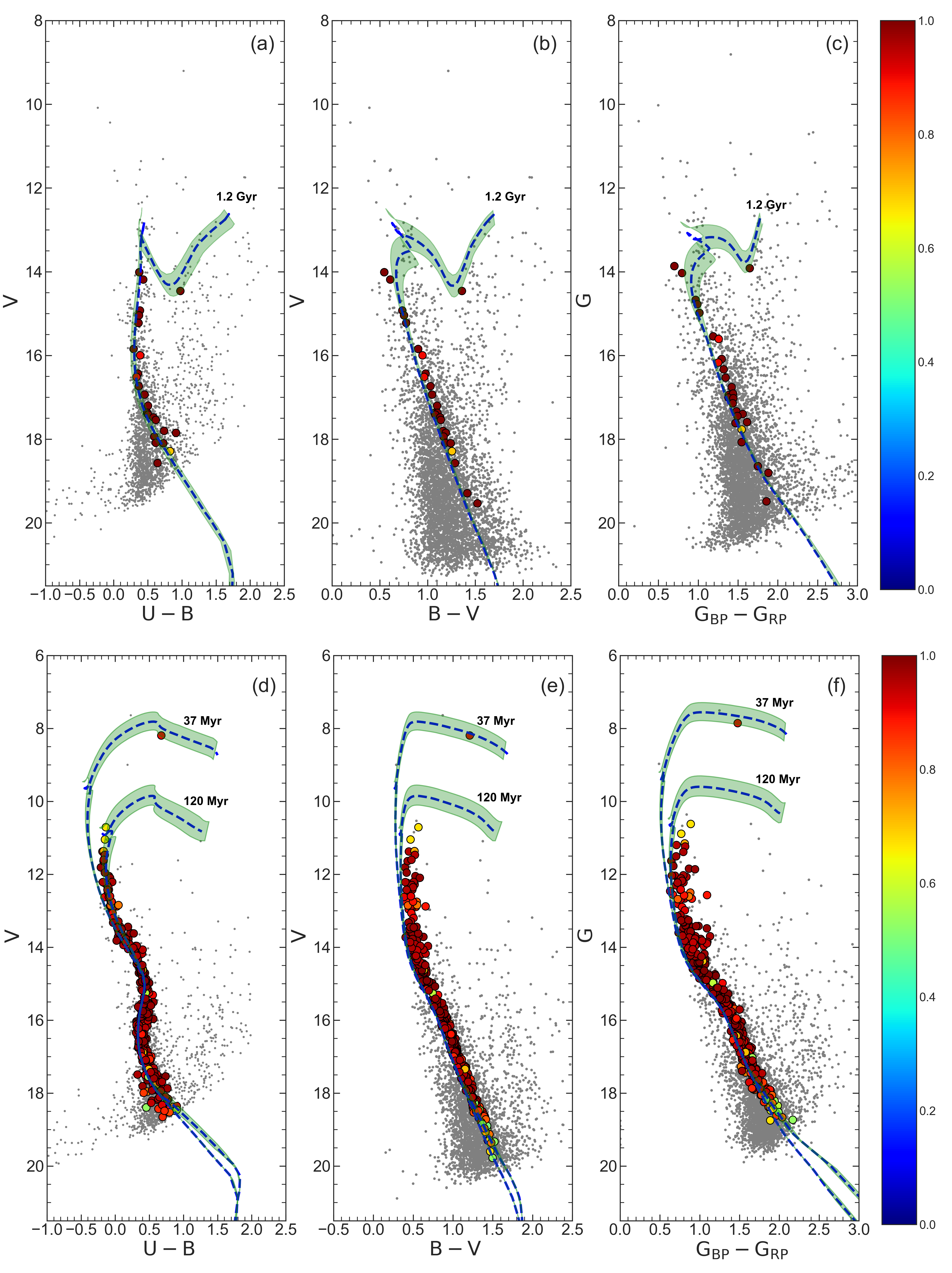}
\caption{$V\times (U-B)$, $V\times (B-V)$ and $G\times (G_{\rm BP}-G_{\rm RP})$ CMDs for the Czernik 2 (sub-figures a, b, and c) and NGC 7654 (sub-figures d, e, and f) clusters. The differently coloured dots represent the membership probabilities according to the colour scales shown on the right side of the diagrams. Grey coloured dots identify the field stars. The blue dashed lines indicate the isochrones (of the ages noted in the diagram). The green shaded areas surrounding these lines are their associated errors. The blue dashed lines in the panels d, e, and f are isochrones corresponding to 37 and 120 Myr ages. The best fitting isochrone for NGC 7654 corresponds to a 120 Myr age for the cluster, while that for Czernik~2 is 1.2 Gyr.} 
\end {figure*}


\subsection{Comparison of Astrometric Results}

Using {\it Gaia} EDR3 astrometric data \citep{Gaia20}, we calculated both mean proper motion components and trigonometric distances from {\it Gaia} for each clusters. Measurement of mean astrometric values was based on the most probable cluster members. To obtain trigonometric distances from {\it Gaia} ($d_{\rm Gaia~EDR3}$), we first converted the trigonometric parallaxes into distances for each member star via the linear expression of $d$(pc) = 1000/$\varpi$ (mas), then we plotted the histograms of distances versus number of stars before finally fitting Gaussian models to the distributions (Fig. 12). 

\begin{figure}[ht]
\centering
\includegraphics[scale=0.57, angle=0]{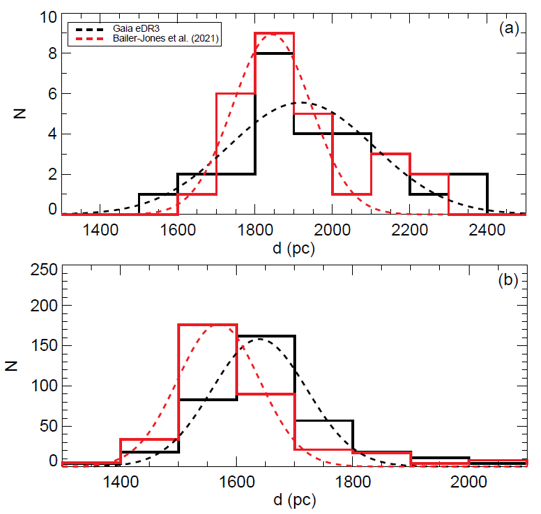}
\caption{The distance histograms of Czernik 2 (a) and NGC 7654 (b). Black and red dashed lines represent the best Gaussian fits to the distances calculated from trigonometric parallaxes \citep{Gaia20} and Bayesian approach method. \citep{Bailer-Jones21}.}
\end {figure}

Just after the second data release of {\it Gaia} \citep{Gaia18}, various researchers deduced numerous values of zero-point offsets for trigonometric parallaxes \citep[i.e.][]{Lindegren18, Riess18, Arenou18}. \citet{Arenou18} estimated several zero-point offsets using parallaxes of different objects (dwarf galaxies, classical variable stars, open and globular clusters). They concluded that zero-point offset values are within the $-0.01$ to $-0.1$ mas depending on the type of object and also are a function of systematic errors, positions on the sky, and magnitude/colour of the objects. In addition to this, there are many zero-point offsets whose values are among the $-0.029$ mas \citep{Lindegren18} and $-0.082$ mas \citep{Stassun18}. These were calculated by taking into account trigonometric parallaxes of quasars, classical cepheids, eclipsing binaries samples, RR Lyrae stars, {\it Kepler} red giant branch and red clump stars \citep{Lindegren18, Stassun18, Riess18, Khan19, Hall19}. Moreover, some researchers reported that the offset correction of parallaxes in {\it Gaia} DR2  may increase for distances in excess of 1 kpc \citep{Stassun18, Lohr18}. The commonly accepted value for the zero-point offset correction was $-0.029$ mas in {\it Gaia} DR2 which was determined from the quasars \citep{Lindegren18}. Recently, some research groups investigated parallax zero-point values from {\it Gaia} EDR3 data taking into account parallaxes of classical cepheids, blue RR Lyrae stars and eclipsing binaries, quasars and red clump stars \citep{Riess21, Bhardwaj20, Lindegren20, Stassun21, Huang21}. According to these studies, the zero-point offset values range from  $-0.037$ mas \citep{Stassun21} to $0.026$ mas \citep{Huang21}. As mentioned in previous studies involving {\it Gaia} DR2 parallaxes corrections, the zero-point offset values are functions of factors such as colour index, magnitude, and position on the sky \citep{Lindegren18}. In this study, we transformed the distances estimated via fitting PARSEC isochrones and used the {\it Gaia} EDR3 data to parallaxes linear expression, comparing the values with each other to understand how these change. For Czernik 2 the isochrone distance ($d_{\rm iso}$) and {\it Gaia} EDR3 distance ($d_{\rm Gaia~EDR3}$) parallaxes are subtending to $\varpi_{\rm iso}=0.531\pm 0.018$ and $\varpi_{\rm Gaia~EDR3}=0.521\pm 0.051$ mas, respectively. The difference between two trigonometric parallax values is $\varpi_{\rm iso} - \varpi_{\rm Gaia~EDR3}=0.010\pm 0.054$ mas. For NGC 7654 the distances correspond to $\varpi_{\rm \rm iso}=0.517\pm 0.039$ and $\varpi_{\rm Gaia~EDR3}=0.610\pm 0.031$ mas from which the difference is $\varpi_{\rm iso} - \varpi_{\rm Gaia~EDR3}=-0.093\pm 0.050$ mas. 

\begin{table*}
\setlength{\tabcolsep}{1pt}
\small{
  \centering
  \caption{Comparison of distances estimated from isochrone fitting and trigonometric parallaxes of {\it Gaia} DR2 ($\varpi_{\rm Gaia~DR2}$) and {\it Gaia} EDR3 ($\varpi_{\rm Gaia~EDR3}$). Here, $\varpi_{\rm iso}$ is trigonometric parallax derived from isochrone distance ($d_{\rm iso}$), $\Delta \varpi_1$ and  $\Delta \varpi_2$ indicate the differences between ($\varpi_{\rm iso}$-$\varpi_{\rm Gaia~DR2}$) and ($\varpi_{\rm iso}$- $\varpi_{\rm Gaia~EDR3}$), respectively.}
\begin{tabular}{lccccccccc}
\hline
Cluster & $d_{\rm iso}$ & $d_{\rm Gaia~DR2}$ & $d_{\rm Gaia~EDR3}$ & $\varpi_{\rm iso}$	& $\varpi_{\rm Gaia~DR2}$	& $\varpi_{\rm Gaia~EDR3}$& $\Delta \varpi_1$ &  $\Delta \varpi_2$ & Reference \\
  & (pc) & 	(pc) & (pc) & (mas)	& (mas)	& (mas) & ($\mu$mas) &  ($\mu$mas) & \\
\hline
ASCC 115      & 732$\pm$69	& 755$\pm$14	& 732$\pm$30   & 1.366$\pm$ 0.128 & 1.325$\pm$ 0.025 & 1.366$\pm$ 0.056 & +41  &   0  & \citet{Yontan19}\\
Collinder 421 & 1245$\pm$103    & 1220$\pm$99	& 1224$\pm$70  & 0.803$\pm$ 0.067 & 0.820$\pm$ 0.066 & 0.817$\pm$ 0.047 & -17  & -14  & \citet{Yontan19}\\
Czernik 2     & 1883$\pm$63	& 1998$\pm$179	& 1919$\pm$189 & 0.531$\pm$ 0.018 & 0.501$\pm$ 0.045 & 0.521$\pm$ 0.051 & +30  & +10  & This study\\
Frolov 1      &	2864$\pm$254    & 2800$\pm$136	& 3098$\pm$275 & 0.349$\pm$ 0.031 & 0.357$\pm$ 0.017 & 0.323$\pm$ 0.029 & -8   & +26  & \citet{Yontan21}\\
Melotte 105   &	2078$\pm$78	& 2460$\pm$180	& 2384$\pm$134 & 0.481$\pm$ 0.018 & 0.407$\pm$ 0.030 & 0.419$\pm$ 0.023 & +74  & +62  & \citet{Banks20}\\
NGC 6793      & 610$\pm$40	& 607$\pm$37	& 590$\pm$34   & 1.639$\pm$ 0.108 & 1.647$\pm$ 0.100 & 1.695$\pm$ 0.098 & -8   & -56  & \citet{Yontan19}\\
NGC 7031      & 1212$\pm$146    & 1404$\pm$81	& 1370$\pm$86  & 0.825$\pm$ 0.099 & 0.712$\pm$ 0.041 & 0.730$\pm$ 0.046 & +113 & +95  & \citet{Yontan19}\\
NGC 7039      & 743$\pm$64	& 767$\pm$41	& 751$\pm$52   & 1.346$\pm$ 0.116 & 1.304$\pm$ 0.069 & 1.332$\pm$ 0.092 & +42  & +14  & \citet{Yontan19}\\
NGC 7086      & 1618$\pm$182    & 1684$\pm$140  & 1673$\pm$116 & 0.618$\pm$ 0.069 & 0.594$\pm$ 0.049 & 0.598$\pm$ 0.041 & +24  & +20  & \citet{Yontan19}\\
NGC 7510      &	2818$\pm$247    & 3450$\pm$477	& 3222$\pm$24  & 0.355$\pm$ 0.031 & 0.290$\pm$ 0.040 & 0.310$\pm$ 0.002 & +65  & +45  & \citet{Yontan21}\\
NGC 7654      & 1935$\pm$146    & 1694$\pm$120	& 1640$\pm$82  & 0.517$\pm$ 0.039 & 0.590$\pm$ 0.042 & 0.610$\pm$ 0.031 & -73  & -93  & This study\\
Roslund 1     & 836$\pm$48	& 883$\pm$54	& 881$\pm$32   & 1.196$\pm$ 0.068 & 1.133$\pm$ 0.069 & 1.135$\pm$ 0.041 & +63  & +61  & \citet{Yontan19}\\
Stock 21      &	1931$\pm$27	& 1934$\pm$159	& 1981$\pm$294 & 0.518$\pm$ 0.007 & 0.517$\pm$ 0.042 & 0.505$\pm$ 0.075 & +1   & +13  & \citet{Yontan19}\\
\hline
    \end{tabular}%
    }
\end{table*}%

\begin{table*}
\setlength{\tabcolsep}{3pt}
  \centering
  \caption{Mean proper motion components and distances ($d_{\rm iso}$, $d_{\rm Gaia~EDR3}$, $d_{\rm BJ21}$) as estimated in this study. The results of \citet[CA20,][]{Cantat-Gaudin_Anders20} and \cite[BJ21,][]{Bailer-Jones21} are also listed in the table.}
\begin{tabular}{l|cccc|ccc|c}
\hline
                  & \multicolumn{4}{c}{This Study} &  \multicolumn{3}{c}{CA20} & BJ21 \\
\hline
     Cluster      & $d_{\rm iso}$ & $d_{\rm Gaia~EDR3}$ & $\mu_{\alpha}\cos \delta$ & $\mu_{\delta}$ & $d$ & $\mu_{\alpha}\cos \delta$ & $\mu_{\delta}$ & $d_{\rm BJ21}$ \\
                  &    (pc)   & (pc) & (mas~yr$^{-1}$)  & (mas~yr$^{-1}$) & (pc) & (mas~yr$^{-1}$) & (mas~yr$^{-1}$) & (pc) \\
\hline
    Czernik 2       &  1883$\pm$63   &  1919$\pm$189  & $-$4.03$\pm$0.04 & $-$0.99$\pm$0.05 & $1907^{+449}_{-305}$   & $-4.06\pm0.01$ & $-0.90\pm0.01$ &  1844$\pm$102 \\
    NGC 7654 &  1935$\pm$146 & 1640$\pm$82  & $-$1.89$\pm$0.03 & $-$1.20$\pm$0.03 & $1600^{+305}_{-221}$ & $-1.94\pm0.01$ & $-1.13\pm 0.01$ & 1569$\pm$71 \\
\hline
    \end{tabular}%
\end{table*}%

We investigated the zero-point offsets for {\it Gaia} trigonometric parallaxes by making a comparison of both the isochrone and the  {\it Gaia} EDR3 distances estimated for the 13 open clusters (also including Czernik 2 and NGC 7654) that we had previously analysed in earlier studies with similar methods \citep{Yontan19, Yontan21, Banks20}. Also, we derived {\it Gaia} DR2 distances of these 13 open clusters and compared the values with both isochrone and {\it Gaia} EDR3 distances to understand how the zero-point offsets change between {\it Gaia} DR2 and EDR3 data. Trigonometric parallaxes corresponding to isochrone and {\it Gaia} EDR3 distances of these open clusters (whose isochrone-based distances are within the range 610 and 2864 pc) were calculated by the linear method. The differences between the methods are listed in Table 7. It can be seen that the median value of the differences in trigonometric parallaxes for {\it Gaia} DR2 and {\it Gaia} EDR3 are 30 and 14 $\mu$mas, respectively, which are an indication of the zero-point offset for the distances of the open clusters. These results show that the uncertainty in the zero-points calculated from {\it Gaia} EDR3 is approximately half of {\it Gaia} DR2. Also, the zero-point offset of 14 $\mu$mas inferred in this study is in the range of values given for {\it Gaia}  EDR3 in the literature for zero-point offsets (see discussion above).

Recently \citet[][BJ21]{Bailer-Jones21} presented a Bayesian approach taking into account a prior with Galactic model parameters, calculating distances with errors for nearly 1.47 billion stars using {\it Gaia} EDR3 trigonometric parallaxes and their uncertainties. We also took into consideration the distances of stars presented by  \citet{Bailer-Jones21} to understand the changes between distances calculated by the trigonometric parallax and Bayesian approaches methods. We retrieved the distance data of the most likely cluster member stars from the catalogue of \citet{Bailer-Jones21} and constructed the histogram of distance versus number of stars (Fig. 12). Then we fitted the distribution with a Gaussian model and determined the mean distances ($d_{\rm BJ21}$) from the data of \citet{Bailer-Jones21}. Errors in distance distribution for both clusters were taken as the standard deviations of Gaussian fits (see Table 8). Two comparisons were made on each histogram as shown in Fig. 12. The distances were estimated as ($d_{\rm Gaia~EDR3}$, $d_{\rm BJ21}$) = ($1919\pm 189,1844\pm 102$) pc and ($d_{\rm Gaia~EDR3}$, $d_{\rm BJ21}$) = ($1640\pm 82,1569\pm 71$) pc for Czernik 2 and NGC 7654, respectively. These results are compatible with those distances derived from isochrone fitting (see Table 8).

 \citet{Cantat-Gaudin_Anders20} estimated mean proper motions and parallaxes of more than 1,200 open clusters using the {\it Gaia} DR2 \citep{Gaia18} astrometric and photometric data, and utilising maximum likelihood procedure. We compared mean proper motion components and distances of the two clusters with those reported by \citet{Cantat-Gaudin_Anders20}. This comparison of mean proper motions and distances ($d_{\rm iso}$, $d_{\rm Gaia~EDR3}$, $d_{\rm BJ21}$) is given in Table 8. For the Czernik 2 we estimated the proper motion components as ($\mu_{\alpha}\cos\delta$, $\mu_{\delta}$) = ($-4.03  \pm 0.04$, $-0.99 \pm 0.05$) mas yr$^{-1}$ and for NGC 7654 as ($\mu_{\alpha}\cos\delta$, $\mu_{\delta}$) = ($-1.89 \pm 0.03$, $-1.20 \pm 0.03$) mas yr$^{-1}$. These are in general agreement with the results of \citet{Cantat-Gaudin_Anders20} and \citet{Liu19}, being inside the $2\sigma$ ranges for both clusters.

\begin{figure}[t]
\centering
\includegraphics[scale=0.7, angle=0]{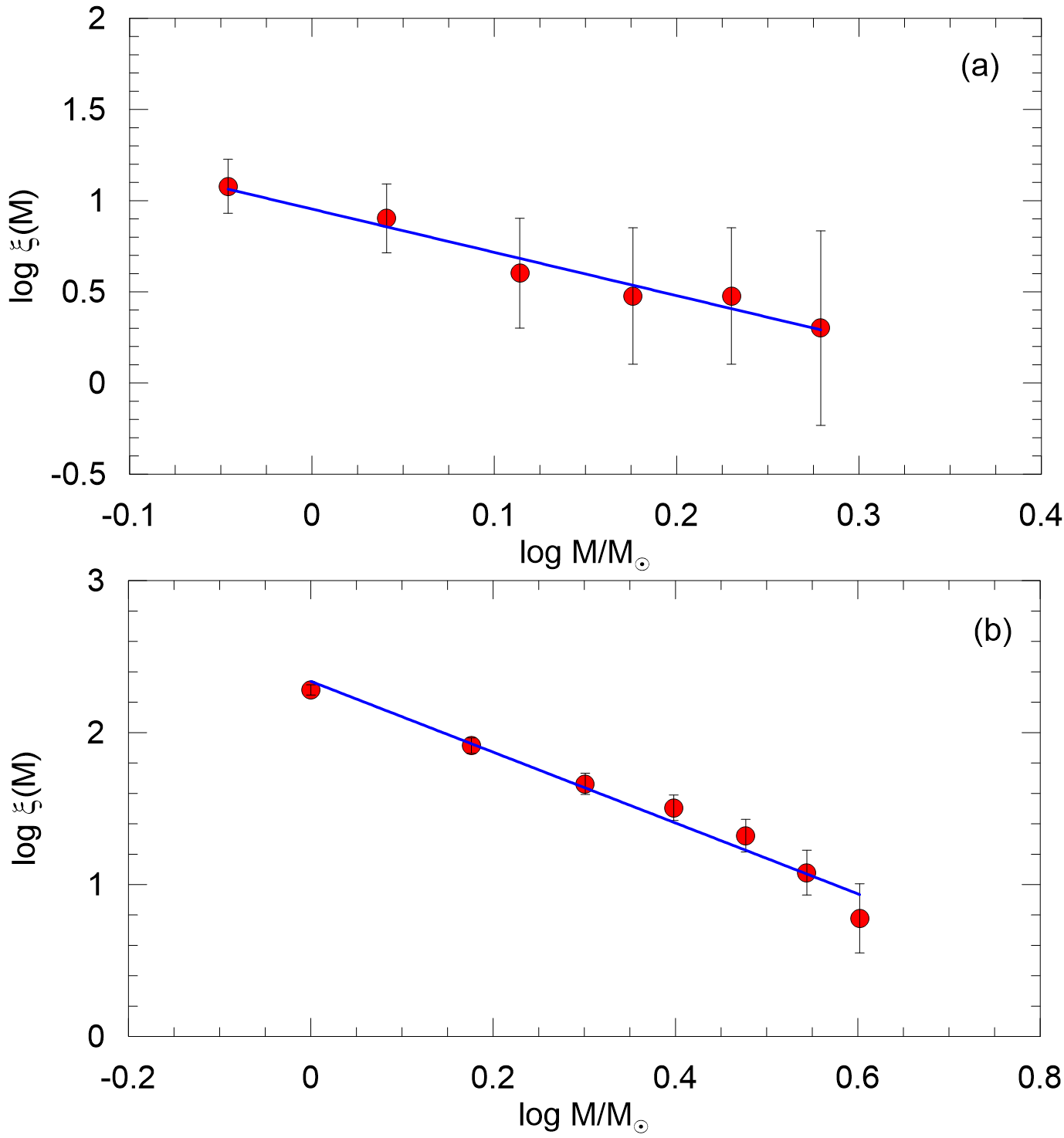}
\caption{Present day mass functions of Czernik 2 (a) and NGC 7654 (b) derived from member stars. Solid blue lines are the mass functions of the two clusters.} 
\end {figure}

\begin{table}
\setlength{\tabcolsep}{3.5pt}
  \centering
  \caption{Present day mass function slopes ($X$) of Czernik 2 and NGC 7654: $N$ is number of stars. The modelled mass range interval is also given.}
    \begin{tabular}{lccc}
\hline
    Cluster & $N$ & $X$ & Mass Range \\
\hline
    Czernik 2 &  32 &  -1.37$\pm$0.24 & 0.80 $<M/M_{\odot}<$ 2 \\
    NGC 7654  & 401 &  -1.39$\pm$0.19 & 0.75 $<M/M_{\odot}<$ 4 \\
\hline
    \end{tabular}
\end{table} 

\subsection{Present Day Mass Functions}
To obtain estimates for the present mass functions of the two clusters we considered 32 and 401 main-sequence stars with probability $P>0$ in Czernik 2 and NGC 7654, respectively. We derived the absolute magnitude of each star using the equation $M_V = V - \mu_V + E(B-V)$ (where $V$ is apparent magnitude of star, $\mu_V$ the distance modulus, and $E(B-V)$ is the colour excess of the cluster). Masses of cluster member stars were calculated from PARSEC models \citep{Bressan12}. A polynomial expression was constructed by correlating a relation between theoretical absolute magnitude and the masses presented in PARSEC models. Using this expression we transformed the absolute magnitudes of stars to their mass values. As a result, we confirmed that the mass ranges of the main sequence stars are lying within $0.80<M/M_{\odot}<2$ and $0.75<M/M_{\odot}<4$ for Czernik 2 and NGC 7654, respectively. To calculate present day mass functions of the clusters, one should first construct the mass distributions. We therefore tabulated the stars within 0.2 and 0.5 $M/M_{\odot}$ mass `steps' for Czernik 2 and NGC 7654, respectively. We identified the number of stars in each mass range and calculated their logarithmic values. Equation (4) was used to determine present day mass functions of two clusters. 
\begin{equation}
\log (dN/dM) =-(1+X)\times \log M + C.   
\end{equation}
$M$ is the mean mass value; $X$ is the slope of the mass function and $C$ expresses the constant of the linear fit. Errors in the mean masses were calculated according to Poisson statistics. The distribution of mass versus logarithmic numbers is shown in Fig. 13, and results are listed in Table 9. 

\section{Conclusion}

In the present paper we derived fundamental astrophysical (reddening, metallicity, distance, and age) and astrometric (mean proper motion components) parameters of Czernik 2 and NGC 7654. These estimates were based on CCD {\em UBV} photometric and {\it Gaia} EDR3 \citep{Gaia20} photometric and astrometric data. 

To obtain cluster membership probabilities we used stellar proper motion components provided by the {\it Gaia} EDR3 catalogue \citep{Gaia20}. However such membership probabilities cannot identify binary star contamination in the main-sequence of a cluster, we therefore also considered photometric criteria to identify real cluster members. To do this, the ZAMS of \citet{Sung13} was fitted on CMDs taking into account the 0.75 mag increase in brightness caused by binary stars. In this way we specified lower and upper bases of the main-sequences for both open clusters. Thus we coupled the membership probabilities with the photometric selection criteria and strengthened the separation of real cluster members. We used those resulting members to determine the fundamental parameters of the clusters.

\begin{enumerate}

\item We estimated structural parameters of the two clusters through fitting the RDP of \citet{King62} (see Table 6). The radii that the cluster stellar densities equal the background stellar density were taken as the physical sizes of clusters, being $r=5'$ and $r=8'$ for Czernik 2 and NGC 7654, respectively.

\item We refined the selection of cluster member stars by taking into account together the membership probabilities, the effect of binary stars, and the physical sizes of the clusters. Thus, for Czernik 2 and NGC 7654 we found 28 and 369 member stars, respectively, with $P\geq 0.5$.

\item Based on the most likely member stars, we estimated the mean proper motions of the Czernik 2 and NGC 7654 as ($\mu_{\alpha}\cos\delta$, $\mu_{\delta}$) = ($-4.03\pm0.04, -0.99\pm 0.05$) mas yr$^{-1}$ and ($\mu_{\alpha}\cos\delta$, $\mu_{\delta}$) = ($-1.89\pm 0.03, -1.20\pm 0.03$) mas yr$^{-1}$, respectively. These values are in very good agreement with the study of \citet{Cantat-Gaudin_Anders20}.

\item By comparing the $(U-B) \times (B-V)$ TCDs with the ZAMS of \citet{Sung13}, we estimated the colour excesses of Czernik 2 and NGC 7654 as $E(B-V)=0.46\pm 0.02$ mag and $E(B-V)=0.57\pm 0.04$ mag, respectively. Both values are consistent with the normal interstellar extinction law. These results are compatible with the study of \citet{Cantat-Gaudin20}. 

\item Photometric metallicities of Czernik 2 and NGC 7654 were based on the calculation of UV excesses of F-G type main-sequence member stars with respect to the main-sequence of Hyades cluster on the TCDs. The metallicites of Czernik 2 and NGC 7654 were estimated as ${\rm [Fe/H]}= -0.08 \pm 0.02$ dex and ${\rm [Fe/H]}= -0.05 \pm 0.01$ dex, respectively. 

\item Keeping constant both reddening and metallicity, we derived distance moduli and ages of two clusters by fitting PARSEC models which have a metallicities $Z=0.013$ for Czernik2 and $Z=0.014$ for NGC 7654 on CMDs. The distance modulus for Czernik 2 was calculated to be $\mu_V=12.80\pm 0.07$ mag and the age $t=1.2\pm 0.2$ Gyr. For NGC 7654 the distance modulus is $\mu_V=13.20\pm 0.16$ mag and its age $t=120\pm 20$ Myr.

\item In the study, we calculated distances using three methods: by fitting isochrones on CMDs, using {\it Gaia} EDR3 trigonometric parallaxes, and distances of member stars which were presented \citet{Bailer-Jones21}. The distances obtained via three methods are in very good agreement within the errors (see Table 8). We can infer that the distance whose determination depends on isochrone fitting gives reliable results. 

\item It has been shown that there is a zero-point offset of 14 $\mu$mas between isochrone distances and {\em Gaia} distances of 13 open clusters examined by homogeneous methods. Moreover, it is seen that the value of 30 $\mu$mas in {\it Gaia} DR2 decreased to 14 $\mu$mas in {\it Gaia} EDR3, which complies with the low zero-point values predicted in {\it Gaia} EDR3.

\item The slopes of the present day mass functions were estimated as $X = -1.37 \pm 0.24$ and $X= -1.39 \pm 0.19$ for Czernik 2 and NGC 7654, respectively. Results are in a good agreement with the value $-1.35$ presented by \citet{Salpeter55}.
\end{enumerate}

\begin{acknowledgements}
The authors are grateful to the anonymous referee for his/her considerable contributions to improve the paper. This study has been supported in part by the Scientific and Technological Research Council (T\"UB\.ITAK) 119F014. We thank T\"UB\.ITAK for partial support towards using the T100 telescope via project 15AT100-738. We also thank the on-duty observers and members of the technical staff at the T\"UB\.ITAK National Observatory for their support before and during the observations. This research made use of VizieR and Simbad databases at CDS, Strasbourg, France. This work has made use of data from the European Space Agency (ESA) mission \emph{Gaia} (https://www.cosmos.esa.int/gaia), processed by the \emph{Gaia} Data Processing and Analysis Consortium (DPAC,  https://www.cosmos.esa.int/web/gaia/dpac/ consortium). Funding for DPAC has been provided by national institutions, in particular the institutions participating in the \emph{Gaia} Multilateral Agreement. IRAF was distributed by the National Optical Astronomy Observatory, which was operated by the Association of Universities for Research in Astronomy (AURA) under a cooperative agreement with the National Science Foundation. PyRAF is a product of the Space Telescope Science Institute, which is operated by AURA for NASA.
\end{acknowledgements}

\label{lastpage}

\end{document}